\documentclass[apj]{emulateapj}

\usepackage[normalem]{ulem}

\usepackage{lineno}

\usepackage{scalerel}
\usepackage{tikz}
\usetikzlibrary{svg.path}

\usepackage{hyperref}

\usepackage{mathtools}
\usepackage{times}

\usepackage{booktabs}

\usepackage[normalem]{ulem}
\usepackage{color}

\definecolor{orcidlogocol}{HTML}{A6CE39}
\tikzset{
  orcidlogo/.pic={
    \fill[orcidlogocol] svg{M256,128c0,70.7-57.3,128-128,128C57.3,256,0,198.7,0,128C0,57.3,57.3,0,128,0C198.7,0,256,57.3,256,128z};
    \fill[white] svg{M86.3,186.2H70.9V79.1h15.4v48.4V186.2z}
                 svg{M108.9,79.1h41.6c39.6,0,57,28.3,57,53.6c0,27.5-21.5,53.6-56.8,53.6h-41.8V79.1z M124.3,172.4h24.5c34.9,0,42.9-26.5,42.9-39.7c0-21.5-13.7-39.7-43.7-39.7h-23.7V172.4z}
                 svg{M88.7,56.8c0,5.5-4.5,10.1-10.1,10.1c-5.6,0-10.1-4.6-10.1-10.1c0-5.6,4.5-10.1,10.1-10.1C84.2,46.7,88.7,51.3,88.7,56.8z};
  }
}

\newcommand\orcidicon[1]{\href{https://orcid.org/#1}{\mbox{\scalerel*{
\begin{tikzpicture}[yscale=-1,transform shape]
\pic{orcidlogo};
\end{tikzpicture}
}{|}}}}

\newcommand{\MS}{M\textsubscript{$\odot$}}    
\newcommand{\RS}{R\textsubscript{$\odot$}}    
\newcommand{\RE}{R\textsubscript{$\oplus$}}   

\newcommand{\degrees}{\ensuremath{^{\circ}}}

\global\dimen\footins = 1.75cm

\colorlet{MYCOLOR}{green!25!red!75!}

\begin{document}

\title{\textcolor{mycolor}{\textsf{NEMESIS}:} Exoplanet tra\textcolor{mycolor}{N}sit surv\textcolor{mycolor}{E}y of nearby \textcolor{mycolor}{M}-dwarfs in t\textcolor{mycolor}{ES}s FF\textcolor{mycolor}{IS} I}

\author{Dax L. Feliz\altaffilmark{1,2,3} \orcidicon{0000-0002-2457-7889} , Peter Plavchan\altaffilmark{4} \orcidicon{0000-0002-8864-1667}, Samantha N. Bianco\altaffilmark{1}\orcidicon{0000-0001-9729-9413} , Mary Jimenez\altaffilmark{4}\orcidicon{0000-0002-5000-9316} , Kevin I. Collins\altaffilmark{4} \orcidicon{0000-0003-2781-3207} , Bryan Villarreal Alvarado\altaffilmark{5}\orcidicon{0000-0002-4788-5147} , Keivan G.\ Stassun\altaffilmark{1,2} \orcidicon{0000-0002-3481-9052}
}

\altaffiltext{1}{Department of Physics and Astronomy, Vanderbilt University, Nashville, TN 37235, USA}
\altaffiltext{2}{Department of Physics, Fisk University, 1000 17th Avenue North, Nashville, TN 37208, USA}
\altaffiltext{3}{Corresponding Author, dax.feliz@vanderbilt.edu}
\altaffiltext{4}{Department of Physics and Astronomy, George Mason University, 4400 University Drive, MSN 3F3, Fairfax, VA 22030, USA}
\altaffiltext{5}{Escuela de Física, Universidad de Costa Rica, 11501-2060 San José, Costa Rica}

\begin{abstract}
In this work, we present an analysis of 33,054 M-dwarf stars, located within 100 parsecs, via the Transiting Exoplanet Survey Satellite (TESS) full-frame images (FFIs) of observed sectors 1–5. We present a new pipeline called \textsf{NEMESIS}, developed to extract detrended photometry, and to perform transit searches of single-sector data in TESS FFIs. As many M-dwarfs are faint, and are not observed with a two-minute cadence by TESS, FFI transit surveys can provide an empirical validation of how many planets are missed, using the 30-minute cadence data. In this work, we detect 183 threshold crossing events, and present 29 candidate planets for sectors 1–5, 24 of which are new detections. Our sample contains orbital periods ranging from 1.25 to 6.84 days, and planetary radii from 1.26 -- 5.31 \RE. With the addition of our new planet candidate detections, along with detections previously observed in sectors 1–5, we calculate an integrated occurrence rate of 2.49 $\pm$ 1.58 planets per star for the period range $\in$ [1,9] days and planet radius range $\in$ [0.5,11] \RE. We project an estimated yield of 122 $\pm$ 11 transit detections of nearby M-dwarfs. Of our new candidates, 23 have signal-to-noise ratios $>$7, transmission spectroscopy metrics $>$38, and emission spectroscopy metrics $>$10. We present all of our data products for our planet candidates via the Filtergraph data visualization service, located at \url{https://filtergraph.com/NEMESIS}.
\end{abstract}


\section{Introduction}\label{sec:intro}
\setcounter{footnote}{0}
The Transiting Exoplanet Survey Satellite (TESS; \citealt{Ricker:2014}) is the first nearly all-sky space-based transit search mission and was was launched on 2018 April 18. Over the course of its two-year mission, TESS has covered $\sim 85 \%$ of the sky, which is split up into 26 observational sectors (13 per hemisphere). The TESS observational field of view covers a 24\degrees$\times$96\degrees region of sky, spanning from the ecliptic equator to an ecliptic pole (6\degrees--96\degrees in ecliptic latitude), with a time baseline of approximately 27 days per each of the 26 sectors per celestial hemisphere. In year one of the mission, the southern hemisphere was observed, with year two observing the northern hemisphere. Each sector is continuously covered by Full Frame Images (FFIs) with a 30-minute cadence and pre-selected stars are observed in smaller image cutouts called Target Pixel Files (TPFs) (also commonly referred to as ``postage stamps"), with a two-minute cadence. All TESS images have a pixel scale of 21$''$ per pixel. Full details of the TESS observational strategy and instrument characteristics can be found in the TESS Instrument Handbook\footnote{https://heasarc.gsfc.nasa.gov/docs/tess/documentation.html}.

One of the primary goals of the TESS mission is to survey over 200,000 stars in order to discover planets with periods $< 10$ days and radii $< 2.5$ \RE, orbiting the brightest stars in the solar neighborhood. FFIs greatly increase the mission’s potential for new transit detections, with some simulations projecting $\sim10^3$ FFI transit detections (\citealt{Barclay:2018}; \citealt{Huang:2018}). Due to the large amount of storage and data processing required for FFIs, not every star in the TESS Input Catalog (TIC, \citealt{Stassun:2018}, \citealt{Stassun:2019}) receives two-minute cadence observations.

The pre-selection of stars with two-minute cadence observations primarily comprises bright, isolated stars; many dimmer stars, including M-dwarfs, are typically excluded, and only are observed via the 30-minute cadence. Interest in M-dwarf stars as hosts for planets has increased in recent years, as the field of exoplanet discovery has grown \citep{Plavchan:2006PhD}.  Due to being smaller, cooler main-sequence stars, M-dwarfs typically have habitable zones (where liquid water can exist on a planet’s surface) at much smaller orbital distances, as compared to more luminous stars. As a result, planets that happen to orbit in these habitable zones will orbit more frequently. Since M-dwarfs comprise roughly 70\% of all stars in the Milky Way \citep{Bochanski:2010}, they offer the best current chance of finding and characterizing habitable planets through sheer numbers and proximity to the Sun. M-dwarf stars also offer a plethora of observational opportunities for exoplanet transit detections. Due to M-dwarfs being small main-sequence stars, small planets are easier to detect via the transit method, as photometric transit depths are larger due to the smaller star-to-planet radius ratios. 

Previous studies of the occurrence rates of short-period planets around Sun-like stars have revealed a bimodal distribution in planetary radii, commonly referred to as the radius valley (\citealt{Fulton:2017}; \citealt{Fulton:2018}; \citealt{Mayo:2018}). The transition between rocky and non-rocky planets around Sun-like and low-mass stars have been found to be dependent on orbital period (\citet{VanEylen:2018}; \citet{Martinez:2019}; \citealt{Wu:2019}; \citealt{Cloutier:2020}). The
slope of the radius valley around Sun-like stars has been measured (\citealt{Martinez:2019}, M19 hereafter) using the California-Kepler Survey (CKS) stellar sample from \citealt{Fulton:2017}, where the authors found their slope to be consistent with model predictions from thermally driven atmospheric mass loss \citep{Lopez:2018}. \citealt{VanEylen:2018} (VE18 hereafter) also measured the planet radius–period slope of the radius valley for their sample of FGK stars, characterized by asteroseismology. Another interpretation for the existence of the radius valley around low-mass stars is that thermally driven mass loss is dependent on stellar mass, and is less efficient overall for mid- to late M-dwarfs, due to gas-poor environments (\citealt{Cloutier:2020}, CM20 hereafter). CM20 also found that for low-mass stars, the transition between rocky planets and non-rocky planets in the radius valley is centered around 1.54~\RE.

The Science Processing Operations Center at NASA Ames (SPOC, \citealt{Jenkins:2016}) which produces calibrated light curves and validation reports for threshold crossing events (TCEs) that pass various tests vetted by humans, also produces TESS Objects of Interest (TOIs) for a subset of targets, observed with two-minute cadences. The MIT Quick Look Pipeline (QLP, \citealt{Huang:2020}) also produces calibrated light curves and validation reports for stars observed in FFIs with TESS magnitudes $<$13.5, and the QLP team also contributes to providing additions to the TOI catalog\footnote{https://tess.mit.edu/toi-releases/ \label{ftn:toi_catalog}}. The DIAmante pipeline \citep{Montalto:2020} has been used to conduct a transit survey that produced 396 Community TESS Objects of Interest (cTOIs)\footnote{https://archive.stsci.edu/hlsp/diamante\label{ftn:DIAmante}} for a subset of FGKM stars
observed in TESS FFIs in sectors 1–13. In this work, we explore the detectability of exoplanets transiting nearby M-dwarf stars, using our custom pipeline, \textsf{NEMESIS}, which is designed to extract photometry, and to detect transits observed in TESS FFIs for TESS sectors 1–5.

In Section \ref{sec:targets}, we describe the TESS observations, together with our criteria for selecting target stars on which to conduct our transit survey. In Section \ref{sec:data_reduction}, we discuss our process of extracting photometric time series from the FFI images. In Section \ref{sec:methods},  we discuss our approach to searching for transiting exoplanets, along with our process for vetting planetary candidates, and testing the sensitivity of our pipeline’s ability to detect transit events. In Section \ref{sec:results},  we present our list of planet candidates, and compare them to other known transiting exoplanets and planet candidates from the TOI and DIAmante catalogs, as well as exploring the ability of our pipeline to detect transits of short-period planets ($<$10 days). In Section \ref{sec:discussion}, we discuss our results, the limitations of our analysis, and areas for improvement in future work. In Section \ref{sec:conclusion}, we discuss our final conclusions.

\section{Target Star Selection}\label{sec:targets}
\begin{figure}[htp]
    \includegraphics[width=\columnwidth,height=0.25\textheight]{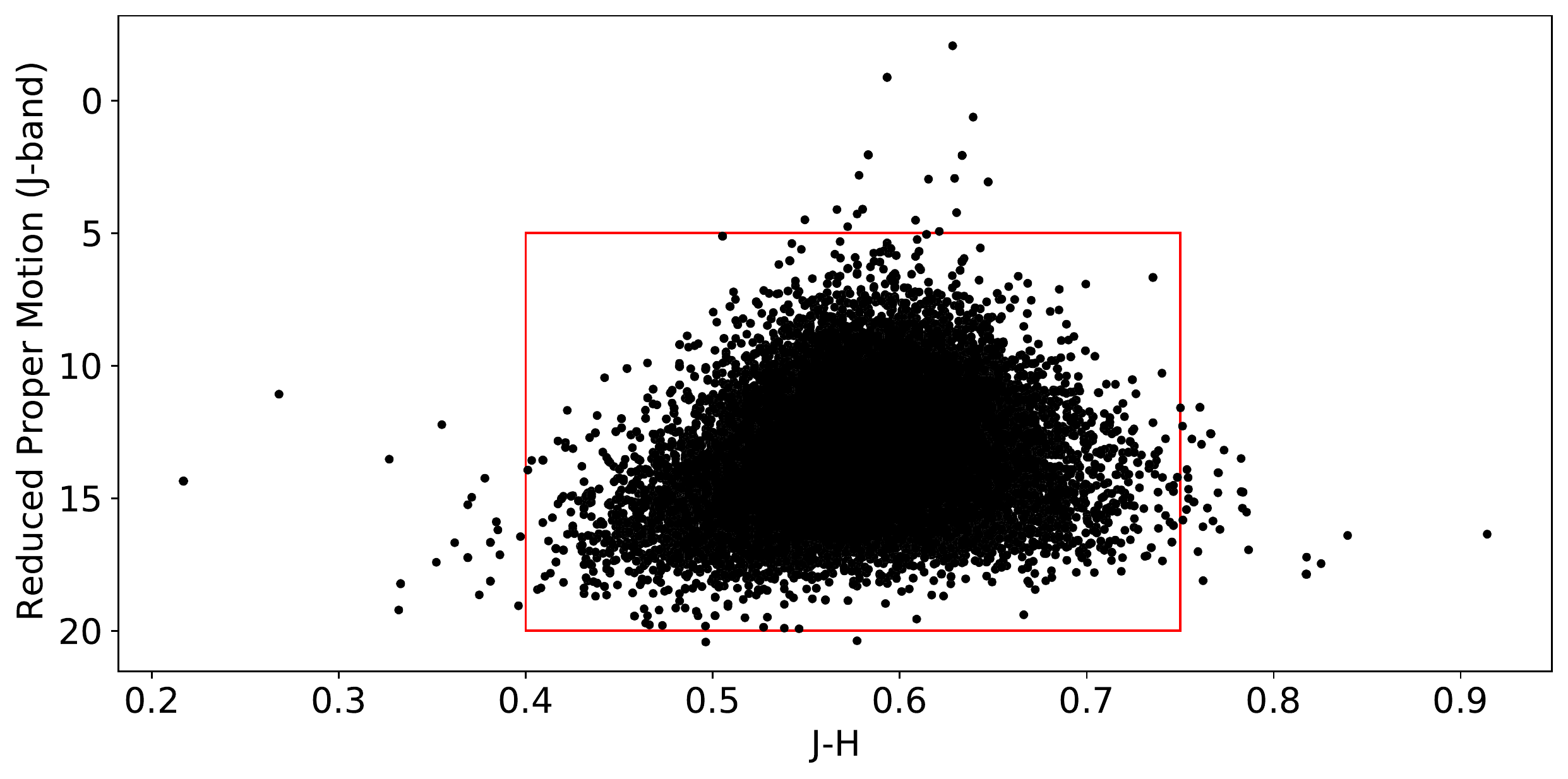}
    \caption{J-H color versus reduced proper motion ($\mathrm{RPM_{J}}$). We selected a J-H cut of 0.4--0.75, and a reduced proper motion cut of 5--20 so as to exclude targets that differed from the majority of the stellar sample. }
    \label{fig:RPM_JmH_cuts}
\end{figure}

In order to filter all the stars observed by TESS, we utilized the TESS Input Catalog Version 8 \citep{Stassun:2019} to access stellar parameters, and the Barbara A. Mikulski Archive for Space Telescopes (MAST) CasJobs\footnote{http://mastweb.stsci.edu/mcasjobs/} service to obtain a target list of M-dwarf stars. The selection criteria used to create our stellar sample is:: $2300~K<\mathrm{T_{eff}}<4000~K$, $\mathrm{TESS~magnitude}<18$, $\mathrm{distance}<100$ pc, $\mathrm{R_{Star}}<0.5~\RS$ and $\mathrm{M_{Star}}<0.5~\MS$. To assist in avoiding contamination from red giants in our target star sample, we also make use of surface gravity cuts (log g $>$ 3). To remove high proper-motion targets, and stars that are bluer or redder than most of our sample stars, we employ a reduced proper-motion cut in the \textit{J}-band of of $5 < \mathrm{RPM_{J}} < 20$ and a color cut $0.4 < \mathrm{J-H} < 0.75$ as shown in Figure \ref{fig:RPM_JmH_cuts}. Here, $\mathrm{RPM_{J}}$ is defined as:

\begin{equation}
    \mathrm{RPM_{J} = J - 5 log(\sqrt{\mu_{\delta}^{2} +\mu_{\alpha}^{2} \cos{\delta} })}
\end{equation}

\begin{figure}[htp]
    \includegraphics[width=\columnwidth]{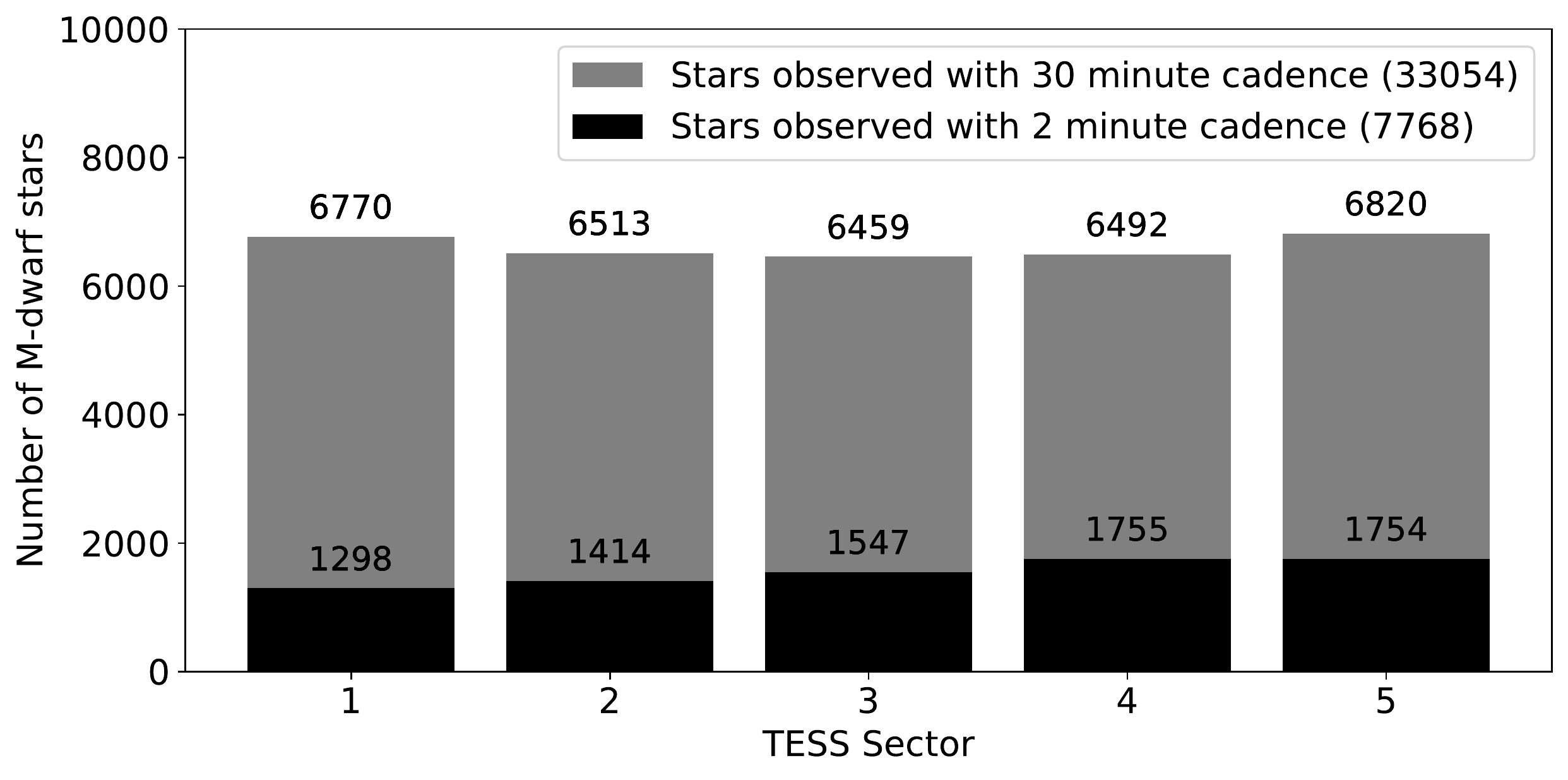}
    \caption{Number of M-dwarf stars observed with two-minute and 30-minute cadences.}
    \label{fig:SPOC_vs_NEMESIS_targets}
\end{figure}

To ensure that our target stars are likely M-dwarfs, we also cross-matched our sample with stars labelled with the “cooldwarf\_v8” flag, as given in the Cool Dwarf Catalog \citep{Muirhead:2018} within the special lists category of the TIC. In Figure \ref{fig:SPOC_vs_NEMESIS_targets}, we display the number of M-dwarf stars meeting our selection criteria, and observed with two-minute and 30-minute cadences, respectively.

\begin{figure*}[htp]
    \includegraphics[width=0.95\textwidth,height=0.3\textheight]{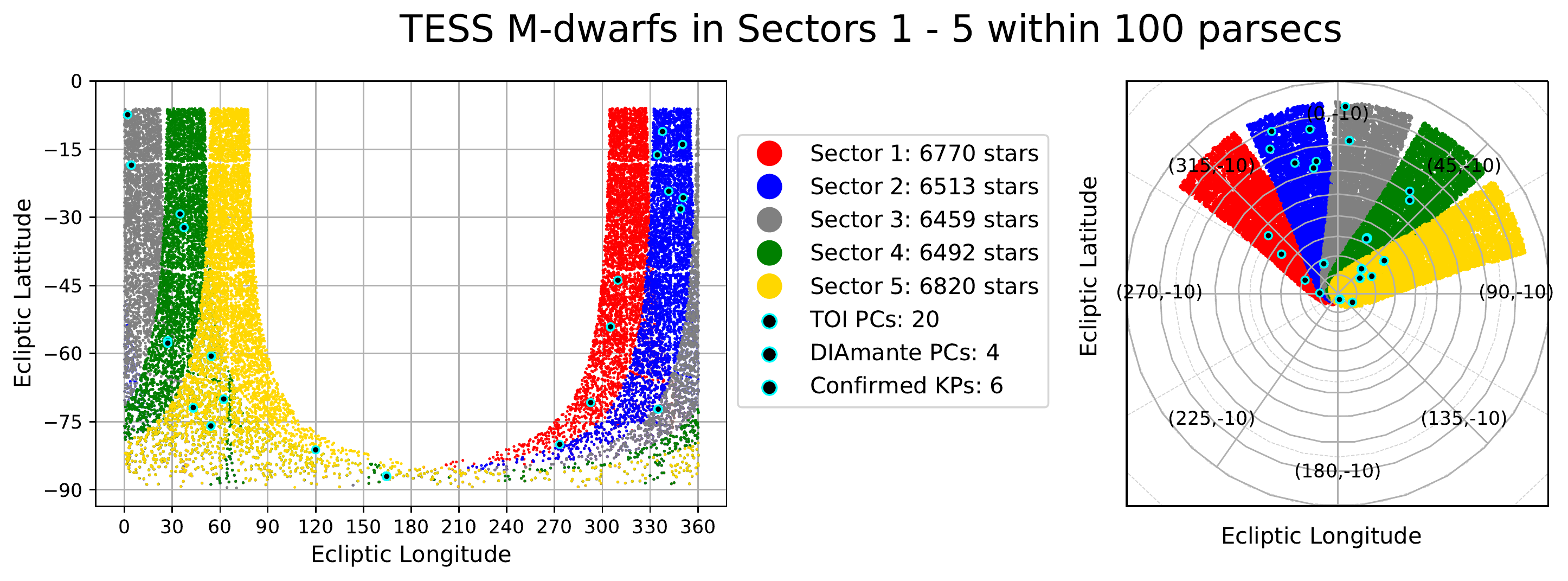}
    \caption{Coordinates of all M-dwarfs in TESS sectors 1--5 within 100 pc. Left: target lists shown in ecliptic coordinates. Right: target lists projected onto celestial sphere, with southern ecliptic pole centered at (0, -90). Known transiting planets observed in sectors 1–5, and confirmed as (KPs), TOI, and DIAmante Planet Candidates (PCs) transiting their M-dwarf host stars, are marked as black points with cyan outlines.}
    \label{fig:target_stars_locations}
\end{figure*}

\begin{figure*}[htp]
    \includegraphics[width=0.95\textwidth, height=0.375\textheight]{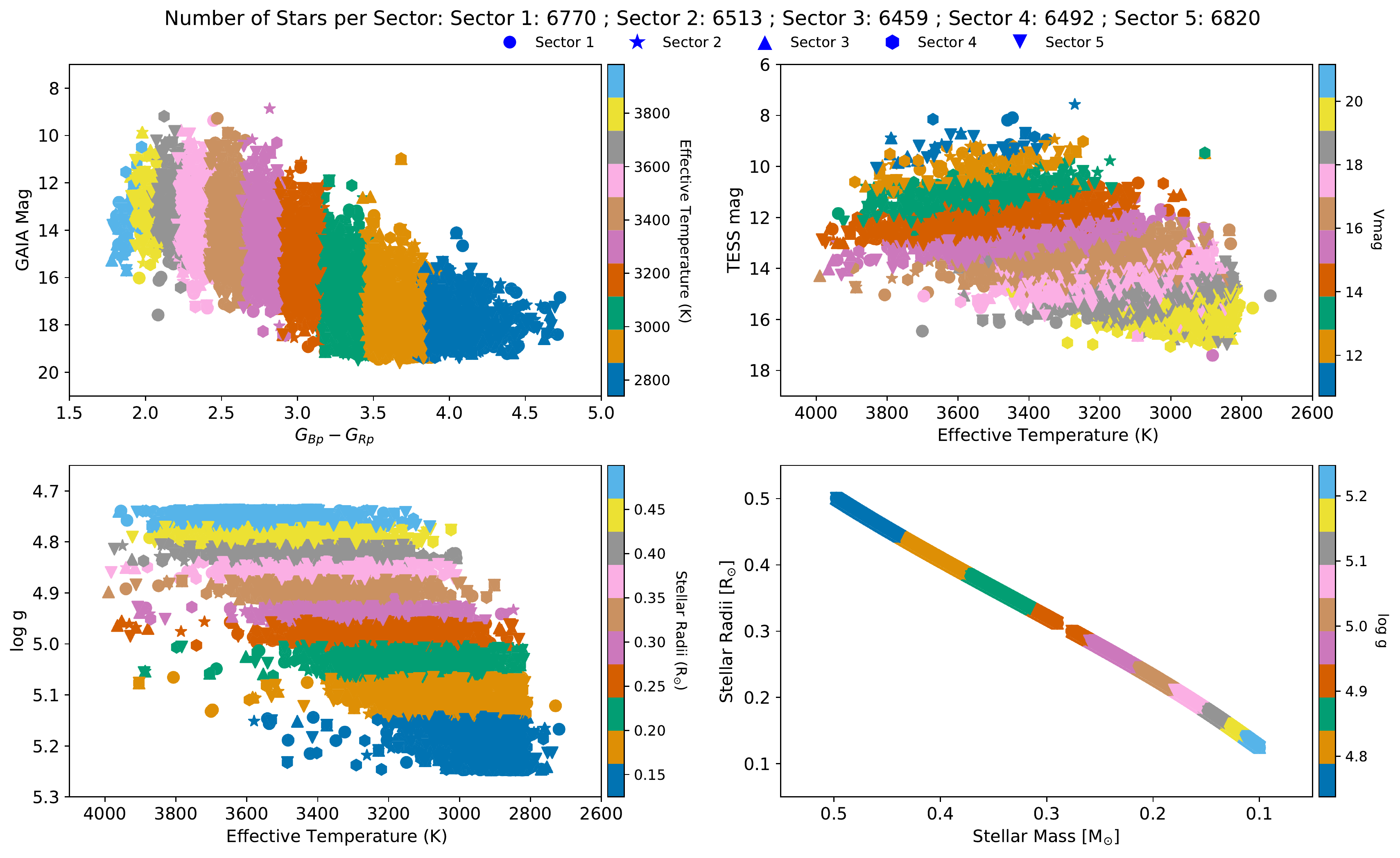}
    \caption{Stellar parameters of all M-dwarfs in TESS sectors 1--5 that follow our selection criteria, as described in Section \ref{sec:targets}.}
    \label{fig:target_star_params}
\end{figure*}

In Figure \ref{fig:target_stars_locations}, we display the ecliptic coordinates of our M-dwarf target stars in TESS sectors 1--5, along with other known planets from the Exoplanet Archive\footnote{https://exoplanetarchive.ipac.caltech.edu/\label{ftn:exoplanet_archive}}, and planet candidates from the TOI and DIAmante catalogs for nearby M-dwarf host stars. In Figure \ref{fig:target_star_params},  we display the distributions of various stellar parameters for our target list.

\section{Data Acquisition and Reduction}\label{sec:data_reduction}
\begin{figure}[htp]
    \includegraphics[width=\columnwidth, trim=0cm 0.0cm 0cm 0.0cm, clip=true]{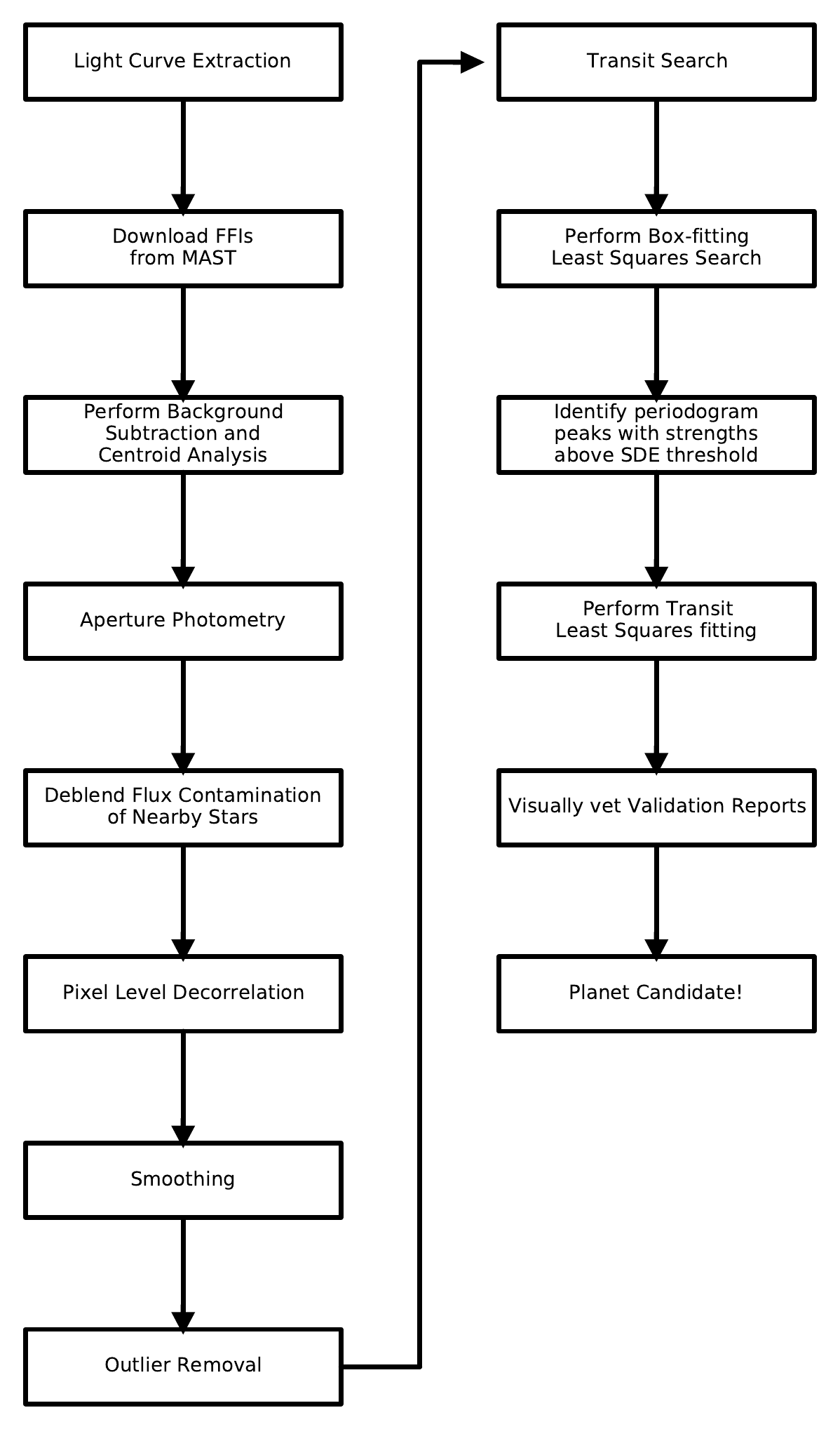}
    \caption{A schematic of the steps in our light curve extraction (left) and transit search pipeline (right). These steps are detailed in Sections 3.1-3.6.}
    \label{fig:flowchart}
\end{figure}

In Sections \ref{sec:FFI_processing} to \ref{sec:Outlier_removal}, we describe our light-curve extraction pipeline, as outlined in Figure \ref{fig:flowchart}. 

\subsection{Processing Full-Frame Images}\label{sec:FFI_processing}

Using the Python package \textsf{Astroquery} \citep{Astroquery:2019}, to query the TIC from MAST, we obtain the stellar radius and mass, which we can then use to cross-match the quadratic limb-darkening parameters from \citet{Claret:2017} for each target star, using the ``catalog\_info" function from the \textsf{Transit Least Squares} \citep{Hippke:2019} package. We then perform a radial cone search for each target star, using a radius of 21\arcsec, and obtain the Right Ascension and Declination coordinates. Having obtained the celestial coordinates, we then download each target star’s FFI square cutout (11 $\times$ 11 pixels in size) from MAST, using the \textsf{TESSCut} \citep{Brasseur:2019} service. 

\subsection{Background Subtraction and Aperture Photometry}\label{sec:SAP_photometry}
TESS periodically fires its thrusters in order to unload the angular momentum built up from solar photon pressure at perigee, and throughout its orbit. These periodic firings are commonly referred to as momentum dumps, and are described in more detail in the Data Release Notes (DRN)\footnote{https://archive.stsci.edu/missions/tess/doc/tess\_drn/} produced for each sector. Moreover, the DRN contain brief descriptions of observations containing a high degree of telescope jitter or glare from Earth and/or the Moon in the FFI field of view. For each FFI cutout around the target star, we first refer to the quality flags described by the DRN to remove specific cadences (with non-zero bitmask flag values) where various types of anomaly are detected. We also remove periods of spacecraft adjustments to maintain pointing stability, also discussed in the DRN.

To estimate the pixel masks for the sky background in each image, we find the dimmest pixels in the FFI cutout, using a brightness threshold of 1/1000 standard deviations below the median flux of the images. We then sum the counts in the background mask, and normalize the background flux. To identify the optimal aperture mask around the target star, we perform a centroid analysis around the target’s pixel position on the median brightness image, using a bivariate quadratic function to approximate the core of a point-spread function. This centroiding technique was also used in the Eleanor FFI pipeline \citep{Feinstein:2019} and is described in more detail
in \citealt{Vakili:2016}.

Having established an approximation of the target star’s location in the FFI cutouts, we then look for pixels displaying flux above a brightness threshold of 7.5 standard deviations above the median flux of the images. In comparison to the SPOC optimal apertures for TPFs, we found that a brightness threshold of 7.5 standard deviations is typically a close approximation for both FFIs and TPFs. To calculate the background-subtracted flux for our selected pixel masks, we use a simple aperture photometry (SAP) approach, defined as

\begin{equation}    
    \mathrm{F_{BKG} = \frac{\sum_{i,j} F(ap_{~BKG}(i,j))}{N_{pix,ap_{~BKG}}}}
\end{equation}

\begin{equation}
\begin{aligned}
    \mathrm{F_{SAP} =  \sum_{i,j}~~F(ap_{~TGT}(i,j)) - F_{BKG}\times N_{pix,ap_{~TGT}}}
\end{aligned}    
\end{equation}
 
where F is the image flux, ap denotes the background (BKG) and target (TGT) pixel masks, i and j are the pixel coordinates of each image, and Npix is the number of pixels in the masks. With the background flux subtracted, we then normalize the SAP flux to have a baseline centered at a value of one. In order to remove sharp changes in flux due to momentum dumps without quality flag values, indicating bad data, we remove data points occurring 30 minutes before and after each thruster firing event.

\subsection{Deblending Flux Contamination from Nearby Stars}\label{sec:flux_contamination}
To account for potential extra light from nearby stars that might contaminate our aperture photometry, we query the TIC for all sources within three TESS pixels of the target star ($\sim$1 arcminute), and record their TESS magnitudes. Typically, our automatically selected apertures spread across 2–3 TESS pixels, so a radial cone search with a width of three TESS pixels is a reasonable general approximation. We calculate the dilution factor as

\begin{equation}
    \mathrm{f = 1 - \frac{10 ^{-T_{mag,target}/2.5}}{ \sum_{i}^{N_{stars~in~aperture}} 10 ^{-T_{mag,i}/2.5}}}
\end{equation}

We then subtract our background-subtracted SAP flux by the dilution factor, and renormalize.

\subsection{Pixel Level Decorrelation}\label{sec:PLD}
Pixel Level Decorrelation (PLD) is an effective technique \citep{Deming:2015, Luger:2016, Luger:2018}, used to remove the effects of intrapixel fluctuations attributed to telescope jitter, or short-term variations such as those caused by momentum dumps, which may occur during TESS observations in each sector. By identifying instrumental systematics with methods such as PLD, the rate of false-positive detections can be decreased for period-searching algorithms such as Box-fitting Least Squares (BLS, \citealt{Kovacs:2002}). Based on Equations (1)--(4) from \citealt{Deming:2015}, we utilize a third-order PLD to calculate a noise model for the observed time
series, so as to model the intrapixel correlations. We then perform a principal component Analysis (PCA) to reduce the number of eigenvectors in our PLD noise model, in order to construct a PCA design matrix and solve for the weights of the PLD model. Having modeled the instrumental systematics, we can then detrend our SAP flux, using the PLD noise model. Based on our testing, we find that using a combination of a third-order PLD with three PCA terms represents the optimal approach to removing short-term intrapixel trends in TESS FFI photometry.

\subsection{Smoothing}\label{sec:Smoothing}
With the instrumental systematics modeled and removed via a PLD noise model, we then remove long-term trends, such as the rotational modulation of starspots, by means of a median-based, time-windowed smoother with an iterative robust location estimator, based on Tukey’s biweight \citep{Tukey:1977} using the \textsf{W\={o}tan} \citep{Wotan} Python package. Removing these trends while keeping the shapes and durations of potential transits intact permits easier detection by period-searching transit-finding algorithms. To determine the optimal window size for the smoothing filter, we calculated the longest transit duration for circular orbits (b=0, i=90\degrees, w=90\degrees, e=0) an Earth-sized planet, transiting every 14 days (minimum of about 2 transits per sector):

\begin{equation} \label{eq:T_dur}
\small
    \mathrm{T_{dur} = \frac{P}{\pi}\arcsin\bigg( (R_{Star}+R_{Planet})(\frac{4\pi^2}{GM_{Star}P^2})^{1/3}\bigg)}
\end{equation} 

We then use three times this transit duration as our
smoothing window, which is calculated for each target star using the stellar radius and mass parameters queried from the TIC. 

\subsection{Outlier Removal}\label{sec:Outlier_removal}
\begin{figure*}[htp!]
    \includegraphics[width=0.95\textwidth]{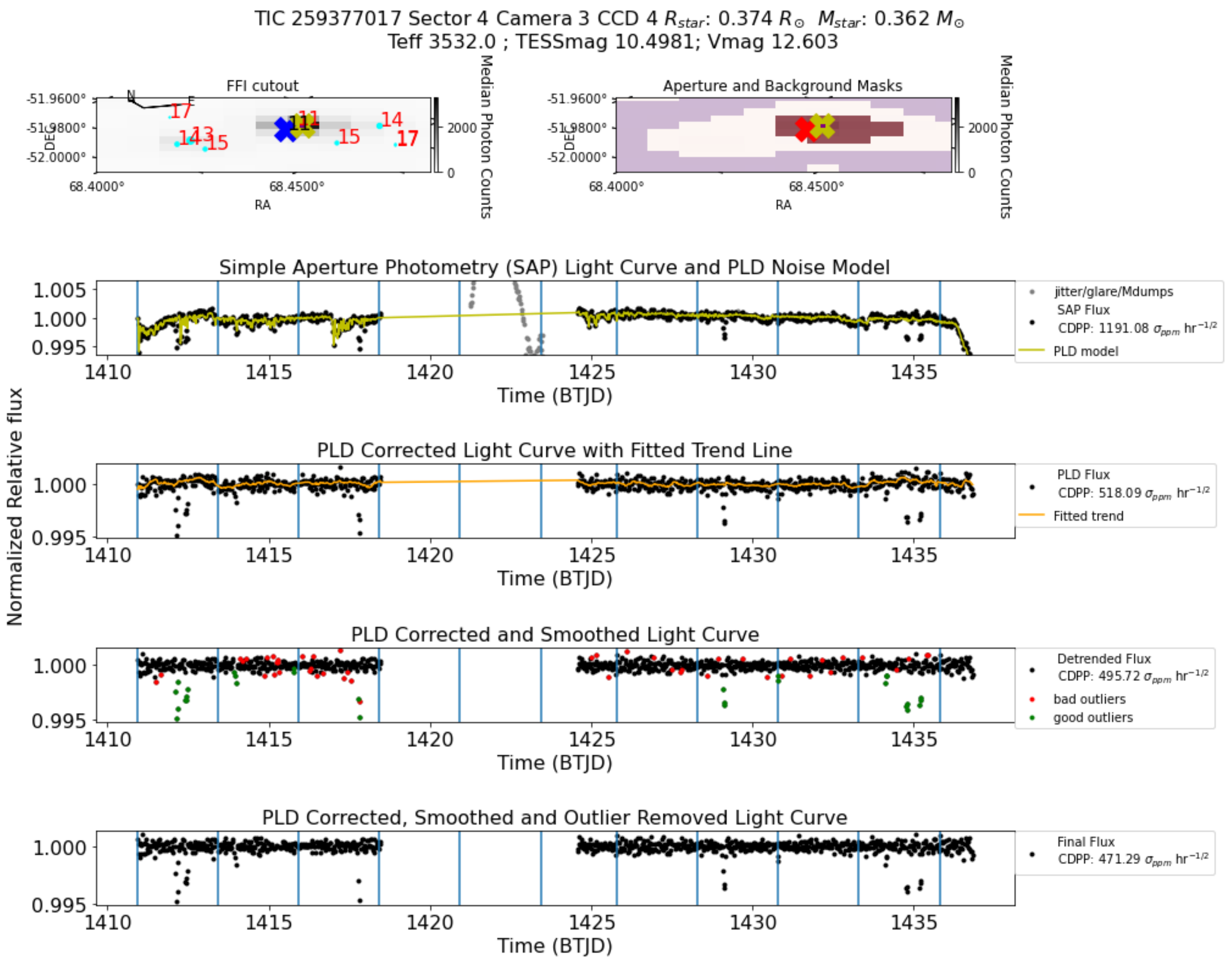}\caption{A step-by-step example of how the NEMESIS pipeline processes a light curve for TIC 259377017, observed in sector 4. Top left panel: FFI cutout with nearby stars and their Gaia magnitudes, marked with cyan points and red text, respectively, on the left panel. The blue X marks the pixel coordinates listed in the FFI headers. The yellow X marks the location of the centroid of the selected aperture. Small purple dots mark the position of the centroid for all images observed in this sector. Top right panel: selected aperture and background masks are marked in red and purple, respectively. The red X here marks the pixel coordinates listed in the FFI headers for visual contrast. Second row: the normalized SAP light curve, indicated by black points. Grey points denote known regions of bad data, as referenced by the DRN (see Section \ref{sec:SAP_photometry}) and data removed around momentum dumps. Momentum dumps are marked by vertical blue lines. The yellow line is the PLD noise model. Third row: the normalized PLD corrected light curve is shown in black points, with the smoothing trend line marked by an orange line. Fourth row: the normalized PLD corrected, smoothed light curve is indicated by black points. Bad outliers that are not consecutive in time are marked in red points. Good outliers that are consecutive in time are marked in green points. Fifth row: final normalized PLD light curve, corrected, smoothed, and with outliers removed. For all processing steps, we display the combined differential photometric precision (CDPP) on the right side of each light curve. These step-by-step summary figures are produced for each star analyzed by NEMESIS.} 
    \label{fig:LC_summary}
    \vspace{0.5cm}
\end{figure*}

Once the SAP flux is extracted, known systematics caused by high jitter and momentum dumps are removed, short-term trends at the pixel level and long-term trends in the time series are removed, and we then begin a sliding-window sigma clipping routine, where we remove outliers higher than $\pm$ 2 median absolute deviations above and below the median flux in a given window of data spanning the same smoothing window estimated via Equation \ref{eq:T_dur}. We are careful to include outliers possessing more than two consecutive data points (“good outliers”) in a given window, so as to not remove potential transit events on $\sim$1 hour timescales. While stellar flares are typically conserved using this approach, for our transit survey, we chose to remove consecutive outliers lying above the local noise thresholds. An example of the full processing of the light curve is displayed in Figure \ref{fig:LC_summary}.

\begin{figure*}[htp!]
    \includegraphics[width=0.95\textwidth]{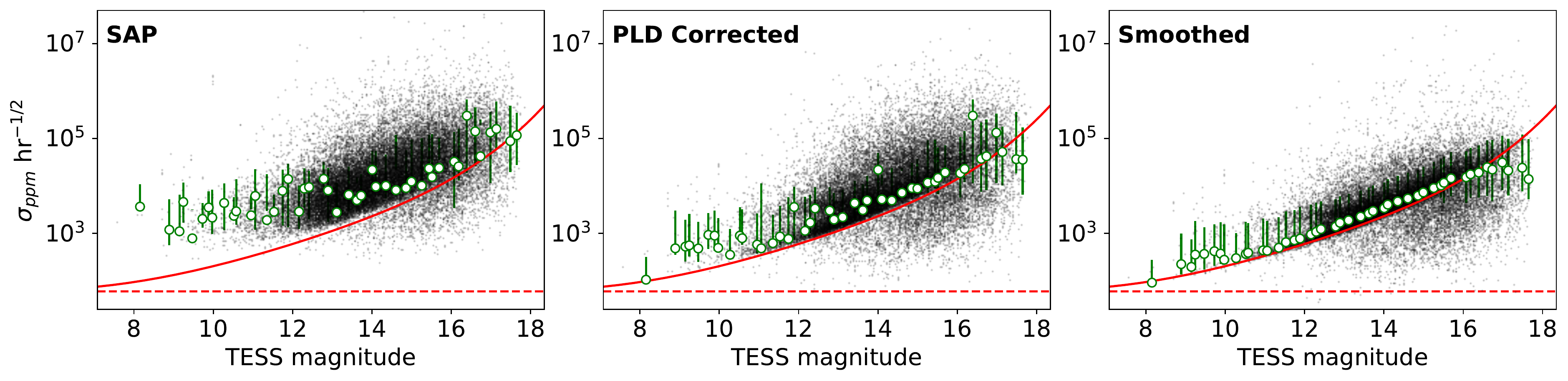}
    \caption{Combined differential photometric precision (CDPP) for all processed light curves from our stellar samples, observed in TESS sectors 1--5, from the beginning of light-curve extraction (left panel) to the end (right panel). The light-curve extraction procedure is described in Section 3. The binned median CDPPs are shown as green circles with the error bars spanning from the 25th quantile to the 75th quantile in each TESS magnitude bin. The solid red line is the expected \citealt{Sullivan:2015} which is a combination of noise contributions from zodiacal light, instrumental read noise, photon-counting noise, and the systematic noise floor. The systematic noise floor (60 ppm hr$^{-1/2}$) is shown as a horizontal dashed red line.}
    \label{fig:cdpp_v_TESSmag_allsectors}
\end{figure*}

To verify the quality of the light curves produced by our pipeline at each step, we calculate the combined differential photometric precision (CDPP; [ppm hr$^{-1/2}$])  as shown in Figure \ref{fig:cdpp_v_TESSmag_allsectors}. We also compare our CDPP values with the preflight
theoretical precision of TESS, estimated \citealt{Sullivan:2015},  and find that the quality of many of our processed light curves exceeds the predicted quality of this model.

\section{Methodology: Transit Detection and Vetting}\label{sec:methods}

\subsection{Transit Searches with BLS and TLS}\label{sec:BLS_TLS}
The Box-fitting Least Squares (BLS) algorithm \citep{Kovacs:2002,Kovacs:2016} is a widely used tool in the exoplanet transit search community. BLS approximates the transit light curve as a boxcar function, with a normalized average out-of-transit flux of one, and with a fixed depth during transit. BLS has a high Signal Detection Efficiency (SDE) for Neptune- and Jupiter-sized planet transits; however, for smaller planets, where the transit depths are comparable to instrumental and stellar noise, the SDE is much lower. This is also a challenge in relation to dim stars, such as M-dwarfs, that may have a lot of photometric scatter.

\begin{figure}[htp]
    \includegraphics[width=0.5\textwidth]{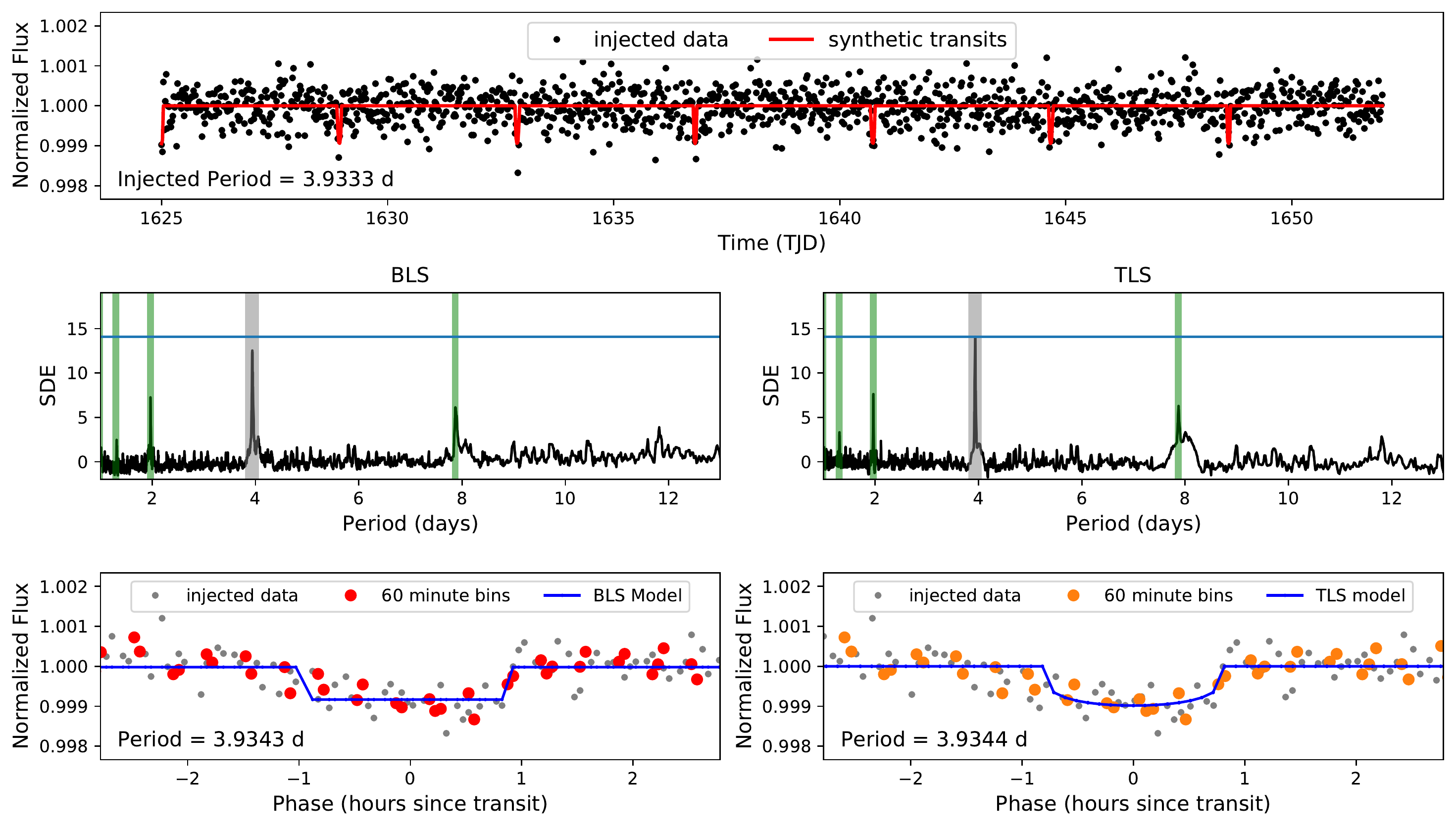}
    \caption{Example recovery of a simulated TESS light curve (30-minute cadence, 400 ppm scatter), injected with a synthetic planet transit (period = 3.93 days, R$\mathrm{_P}$ = 1.5 \RE, circular orbit), using the BLS and TLS methods, respectively. The vertical gray shaded line is the injected period, and the vertical green shaded lines are aliases of the injected period. While the transits are detected by both BLS and TLS, TLS proves to be the more favorable transit-detection algorithm, due to higher SDE measured for its fitted transit models, as compared to the fitted box models produced by the BLS method.}
    \label{fig:BLS_vs_TLS}
\end{figure}

The Transit Least Squares algorithm (TLS\footnote{https://transitleastsquares.readthedocs.io/en/latest/index.html}, \citealt{Hippke:2019}) was created as an alternative analytical approach to the BLS algorithm, in order to be more sensitive to smaller Earth-sized planetary transits. Unlike BLS, which searches for box-like periodic flux decreases in the light curve, TLS utilizes an analytical transit model, with stellar limb-darkening (\citealt{Manduca:1977}; \citealt{MandelAgol:2002}). \citealt{Hippke:2019} found that the false-positive rate was suppressed when comparing algorithms for transit-injected light curves, due to the detection of higher signal-to-noise ratio (S/N). A comparison between the two algorithms on a simulated light curve is displayed in Figure \ref{fig:BLS_vs_TLS}. In this work, we utilize the \textsf{Astropy} implementation of BLS\footnote{https://docs.astropy.org/en/stable/timeseries/bls.html} alongside with TLS.

While TLS may be the optimal method for detecting smaller planets, it is more computationally expensive than BLS to run. To minimize the computation time of transit searches in this survey, we first perform a BLS search, and look for the strongest peaks in the resulting BLS power spectra. In order to select the grid of trial periods to be searched for each target star, we utilize the TLS ``\textit{period\_grid}'' function, which estimates optimal frequency sampling as a function of stellar mass and radius, as detailed in \citet{Ofir:2014}.  Based on this period grid, we
also use the TLS ``\textit{duration\_grid}'' function to estimate the trial fractional transit durations (duty cycles) to fit various transit models over. For the period grid, we use an oversampling factor of 9, and for the duration grid, we use a logarithmic step size of 1.05. In order to avoid false-positive detections due to fast stellar rotations of any active M-dwarf stars with residual trends that may have survived our detrending process, we set the minimum period of our period grid to be one day. We also require each transit search have at least three modeled events, which places an upper limit on the maximum period of our period grid of $\sim$9 days, i.e., $\sim$1/3 of the duration of a single-sector TESS light curve.

Other studies have found empirical SDE thresholds ranging from 6 to 10 \citep{Siverd:2012, Dressing:2015, Pope:2016, Aigrain:2016, Livingston:2018, Wells:2018} to be optimal in reducing the overall number of false positives. While higher SDE thresholds typically result in fewer false positives, lower SDE thresholds provide greater completeness. With this in mind, we utilize a minimum TLS SDE threshold of 10 in this work, in order to reduce the number of potential false-positive detections encountered in our search. When running the faster BLS transit searches, for any target stars displaying strong $\mathrm{SDEs} \geq 6$ in the BLS power spectra, we then run a TLS transit search on the same light curve in order to obtain a better transit fit of the signal causing the initial detection. For TLS transit searches, we use the same period and duration grids used for BLS. For both BLS and TLS transit searches, we record transit parameters corresponding to the strongest peak in their respective power spectra.

\subsection{Vetting Process}\label{sec:vetting}
One of the many challenges of transit surveys is the
weeding-out of false-positive detections by means of a post-detection vetting process. Scenarios that may trigger a false-positive detection include short-term periodic stellar variations, eclipsing binary stars, blended photometry from two stars on one pixel, or periodic movement of the centroid. For our vetting process, we use a one-page validation report (see Appendix~\ref{sec:appendix_validation_reports}),  similar to the validation reports from SPOC, QLP, and DIAmante, containing visual comparisons to other catalogs and archival images (e.g., Gaia, Digitized Sky Survey), as well as several tests to attribute periodic transit-like behavior to planetary transit events. With the metrics produced by these tests, we then visually vet each candidate and refine our TLS searches with finer period and duration grid step sizes. Additionally, we also perform an alternative Lomb–Scargle analysis on each SAP, PLD, and detrended light curve to search for stellar activity that may be triggering transit detections from BLS or TLS. For each TCE, we then vote to determine whether or not it planetary transits is detected, as shown below. 

\subsubsection{Odd/Even Mismatch Test} 
To search for potential transit detections that may be attributed to eclipsing binaries, we utilize the odd–even mismatch test, which compares primary and secondary transit events in orbital phase. For an unequal size ratio eclipsing binary, the secondary event will be a smaller fraction of the primary event’s transit depth, and if the orbit of the transiting star is circular, it will occur at an orbital phase = 0.5. To perform this test, we cut one-day regions around the transit times modeled by TLS, and append these separately into odd- and even-numbered transits. To produce a metric that compares the odd and even transits, we use

\begin{equation}
\mathrm{Odd~Even~Mismatch = \frac{|\delta_{odd} - \delta_{even}|}{\sigma_{\delta_{odd}}+\sigma_{\delta_{even}}}}
\end{equation}
where $\delta$ and $\sigma_{\delta}$ denote the transit depth and transit depth uncertainty of the odd- and even-numbered transits, as measured by TLS. In addition to the odd-even mismatch statistic, we also visually inspect the phase-folded light curves at 0.5, 1 and 2 times the TLS-detected period.

\subsubsection{Centroid Motion Test}
A further challenge when performing automated aperture photometry in an all-sky survey lies in ensuring that the selected apertures remain on-target throughout the duration of transit events. As described in Section \ref{sec:SAP_photometry}, we utilize a bivariate quadratic function to estimate the photocenter around a target star’s pixel position for the image, corresponding to the median brightness. To verify whether or not there is motion of the centroid during the time of transit detected by TLS, we phase-fold the pixel positions of the centroid over time. We track centroid motion by calculating changes in pixel position in the image columns and rows, and subtracting their median pixel positions. If the centroid motion exceeds five standard deviations from the pixel position in the image columns and rows during the time of transit, we consider this a false-positive detection. 

\subsection{EDI-Vetter Unplugged}\label{sec:EDI}
To help provide an alternative reference analysis for our false-positive tests, we utilize a modified version of the \textsf{EDI-Vetter} Python package \citep{Zink:2020}. The \textsf{EDI-Vetter} tool was used to automatically vet planet candidates in campaign 5 of the Kepler K2 mission. The \textsf{EDI-Vetter Unplugged}\footnote{https://github.com/jonzink/EDI\_Vetter\_unplugged \label{ftn:edi_vetter_unplugged}} package is a simplified version of the base package, modified to use the output from TLS. The various false-positive tests performed by the \textsf{EDI-Vetter Unplugged} package are
described in more detail in Section 3 \citealt{Zink:2020} and are briefly summarized here as follows.

\begin{itemize}
    \item[--] \textit{Flux contamination test:} a calculation performed to check for significant contributions of flux from stars that might lie on neighboring pixels.
    \item[--] \textit{Outlier detection test:} a check for outliers during the transit time that might give rise to a false-positive detection.
    \item[--] \textit{Individual transit test:} a comparative check of the S/Ns of individual transits with an SDE threshold $>$10 used in our transit search.
    \item[--] \textit{Even/odd mismatch test:} a check to verify whether the odd–even mismatch metric output by TLS is greater than 5$\sigma$.
    \item[--] \textit{Uniqueness test:} this test identifies cases where the phase-folded light curve appears to contain several transit-like dips that may appear similar to the candidate transits.
    \item[--] \textit{Secondary eclipse test:}a check for statistically significant secondary transit events at 1/2 times the TLS-detected period.
    \item[--] \textit{Phase coverage test:} with the removal of all data relating to momentum-dump periods and bad-quality flags, it is possible for the phase-folded data to contain large gaps in its transit signal. This test is designed to determine whether the transit detection lacks sufficient data to detect a statistically significant transit event.
    \item[--] \textit{Transit duration limit test:} a check to determine whether the detected transit duration is too long for the detected transit period.
    \item[--] \textit{False positive flag:} : if any of the tests listed above listed are flagged as being true, the candidate is flagged as being a potential false positive, requiring closer inspection.
\end{itemize}
 
\section{Results}\label{sec:results}
In this section, we present the results of our search for new transiting planet candidates among M-dwarf stars observed in the TESS FFIs, as well as numerical and modeling tests, performed to assess the detection sensitivity and survey completeness of our detections. A listing of previously identified candidates, and notes relating to whether we confirm or miss those candidates, is provided in Appendix~\ref{sec:appendix_comparison}. Based on our survey completeness and list of planet candidates, we also make an estimate of planet occurrence rates for M-dwarf planet hosts.

\subsection{New Transiting Planet Candidates}
Our transit survey of 33,054 stars observed across TESS sectors 1--5 revealed 183 TCEs with TLS SDEs $\geq$ 10. Of those 183 TCEs, 29 are planet candidates fulfilling our automated search and vetting criteria, 24 of which are new detections. Of our 29 planet candidates, five are detected transits of TOIs 269.01, 270.03 (sectors four and five), 393.01, 455.01, and 1201.01. For all of our PCs, we refined the orbital parameters describing these transiting systems with an MCMC analysis, as described in Section \ref{sec:MCMC}. The best-fit TLS and median posterior MCMC transit orbital parameters (P, T0, $\textrm R_\mathrm{P}$ and b) for each of our planet candidates are displayed in Table \ref{tbl:MCMC_results}.

\begin{figure*}[htp]
    \includegraphics[width=0.95\textwidth]{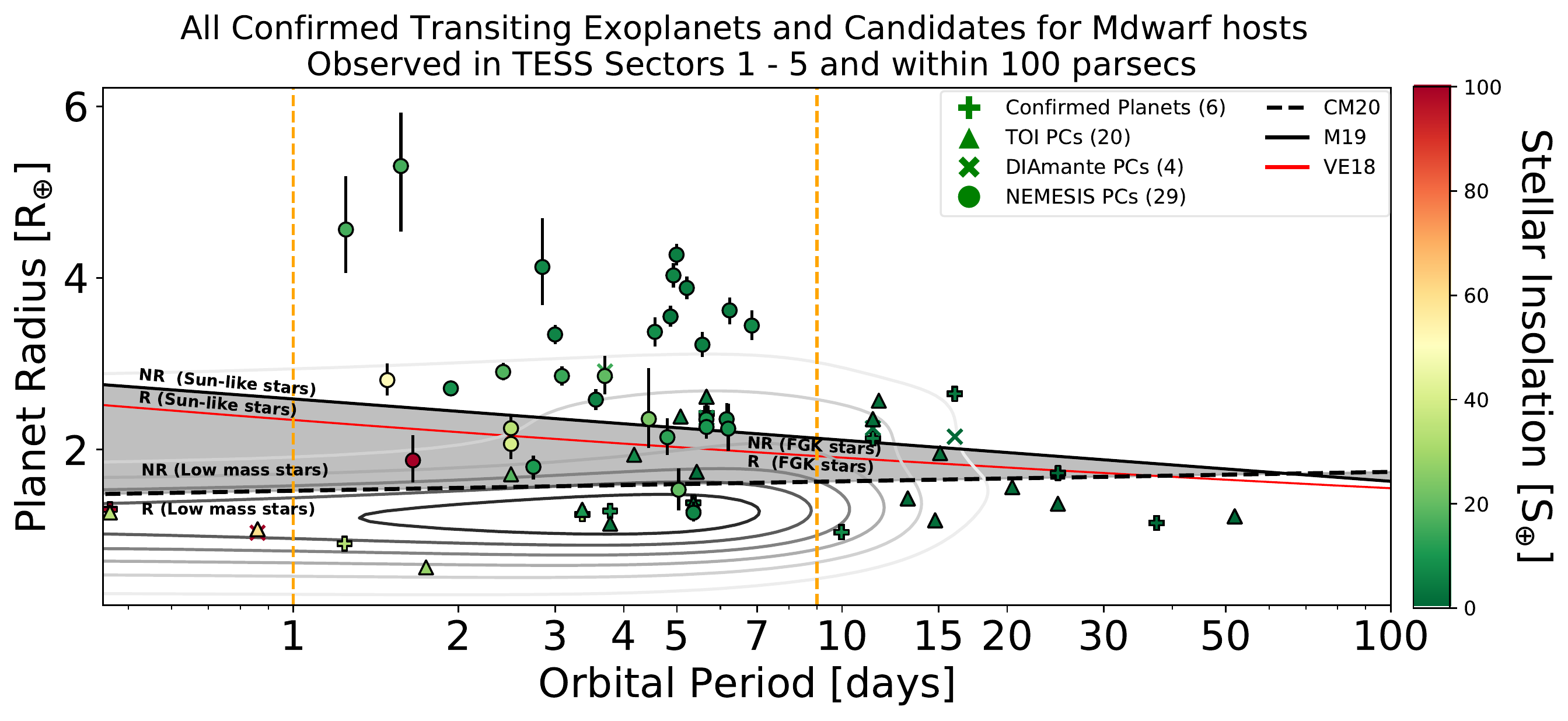}
    \caption{Period-radius diagram of all confirmed transiting exoplanets and exoplanet candidates observed in TESS sectors 1–5. Confirmed transiting planets ($+$s) were queried from the Exoplanet Archive (see footnote ${\ref{ftn:exoplanet_archive}}$) and the TOI planet candidates ($\triangle$s) were queried from the TOI catalog (see footnote ${\ref{ftn:toi_catalog}}$). The DIAmante planet candidates ($\times$s) were queried from the MAST portal (see footnote ${\ref{ftn:DIAmante}}$). Planet candidates from this work are marked as $\bigcirc$s,  and the vertical dashed orange lines represent the range of periods searched by our survey. All points are colored based on their respective incident stellar flux (in Earth units), calculated using Equation \ref{eq:S_insol}. For comparison, we overplot the radius valley slope for FGK stars characterized by asteroseismology (red line; \citealt{VanEylen:2018}, VE18), the approximate period-dependent radius valley slope of Sun-like stars from the CKS (solid black line; \citealt{Martinez:2019}, M19), and the approximate period-dependent radius valley slope of low-mass stars (dashed solid line; \citealt{Cloutier:2020}, CM20). The shaded region highlights the parameter space between the two model estimations of the radius valley from M19 and CM20. Regions above and below these slopes are inferred to be rocky (R) or non-rocky (NR) type planets. The contour lines show the Kernel density estimation of the period–radius space for all planet-hosting M-dwarf systems from the Exoplanet Archive.}
    \label{fig:pc_demographics}
\end{figure*}

To provide context for our planet candidates in regard to planet demographics and the radius valley, we display our planet candidates in the planet radius–period space, and compare them with nearby candidates from the TOI and DIAmante catalogs, together with additional confirmed transiting planets observed in sectors 1--5, as shown in Figure \ref{fig:pc_demographics}. Confirmed transiting planets (marked as $+$s) were queried from the Exoplanet Archive, and the TOI planet candidates (marked as $\triangle$s) were queried from the TOI catalog. The DIAmante planet candidates (marked as $\times$s) were queried from the MAST
portal (see footnote$^{\ref{ftn:DIAmante}}$). Planet candidates from this work are marked as $\bigcirc$s. In addition, we highlight the planet radius–period slopes of the radius valley measured from M19 (for Sun-like stars), VE18 (for FGK stars) and CM20 (for low mass stars). We also color each confirmed planet and planet candidate based on incident stellar flux (S), defined in Earth units $S_{\oplus}$ as

\begin{equation}\label{eq:S_insol}
\mathrm{\frac{S}{S_{\oplus}} = \bigg(\frac{\mathrm{R_{Star}}}{R_{\odot}}\bigg) \bigg(\frac{\mathrm{T_{eff}}}{5777~K}\bigg)^4 \bigg( \frac{1~AU}{ (\frac{\mathrm{M_{star}~G~P^2}}{M_{\odot}~4\pi^2})^{\frac{1}{3}} }\bigg)^2}
\end{equation}

In Figure \ref{fig:pc_demographics}, we also display the Kernel density estimate of the planet radius–period space for all planet-hosting M-dwarf stars from the Exoplanet Archive, using the following criteria: $\mathrm{T_{eff}}<4000 K$, $\mathrm{R_{Star}}< 0.5\RS$, $\mathrm{M_{Star}}<0.5 \MS$ and $\mathrm{logg}>3$. In Appendex \ref{sec:appendix_comparison}, we present a gallery of the phase-folded light curves for our detected planet candidates. Each panel contains the detrended and phase-folded light curve (in black points), and the median transit model obtained via MCMC analysis (red line), as described in Section \ref{sec:MCMC}. Our planet candidates have orbital periods ranging from 1.25–6.84 days, and planetary radii ranging from 1.26 -- 5.31 \RE. In terms of relative incident stellar flux, our candidates range from 3.82 -- 116.18 $\mathrm{S_{\oplus}}$ which is likely to preclude any of our candidates orbiting within the habitable zones of their host stars.

\begin{table*}
\centering
\caption{BEST-FIT AND MEDIAN POSTERIOR SOLUTIONS FOR PLANET CANDIDATE TRANSIT MODELS FROM TLS AND MCMC.}
\label{tbl:MCMC_results}
\resizebox{0.95\textwidth}{!}{\begin{tabular}{rrlllllll}
\toprule
    TIC ID &  TESS Sector &            TLS Period &    TLS Planet Radius & TLS Transit Time &                   MCMC Period &            MCMC Planet Radius &                MCMC Transit Time &         MCMC Impact Parameter \\
\midrule
 139456051 &            1 &   2.8424 $\pm$ 0.0019 &  3.2028 $\pm$ 0.3378 &        1325.3616 &  2.8424$_{-0.0056}^{+0.0056}$ &  4.1286$_{-0.4470}^{+0.5690}$ &  1325.3617$_{-0.0291}^{+0.0296}$ &  0.8256$_{-0.0605}^{+0.1036}$ \\
 143996634 &            1 &   4.9268 $\pm$ 0.0110 &  3.7148 $\pm$ 0.6298 &         1327.406 &   4.927$_{-0.0523}^{+0.0536}$ &  4.0301$_{-0.1411}^{+0.1455}$ &  1327.4058$_{-0.0426}^{+0.0433}$ &  0.7067$_{-0.0298}^{+0.0270}$ \\
 234299676 &            1 &   6.8442 $\pm$ 0.0159 &  2.9942 $\pm$ 0.7537 &         1329.841 &  6.8449$_{-0.1073}^{+0.1084}$ &  3.4455$_{-0.1686}^{+0.1775}$ &  1329.8405$_{-0.0583}^{+0.0584}$ &  0.7169$_{-0.0469}^{+0.0380}$ \\
 281461138 &            1 &   5.2113 $\pm$ 0.0131 &  3.8348 $\pm$ 0.6506 &        1325.5449 &  5.2109$_{-0.0677}^{+0.0688}$ &  3.8846$_{-0.1309}^{+0.1340}$ &  1325.5453$_{-0.0450}^{+0.0445}$ &  0.7433$_{-0.0263}^{+0.0225}$ \\
 290048573 &            1 &   1.4829 $\pm$ 0.0030 &  2.1006 $\pm$ 0.1330 &        1326.6596 &  1.4829$_{-0.0045}^{+0.0044}$ &  2.8102$_{-0.1799}^{+0.1931}$ &  1326.6598$_{-0.0333}^{+0.0334}$ &  0.7044$_{-0.0708}^{+0.0535}$ \\
 389662862 &            1 &   2.9981 $\pm$ 0.0044 &  3.2191 $\pm$ 0.3366 &        1325.9514 &   2.998$_{-0.0131}^{+0.0132}$ &   3.342$_{-0.1143}^{+0.1122}$ &  1325.9516$_{-0.0352}^{+0.0353}$ &  0.6259$_{-0.0411}^{+0.0328}$ \\
   9632592 &            2 &   4.9924 $\pm$ 0.0087 &  3.8078 $\pm$ 0.6487 &        1355.3707 &  4.9921$_{-0.0426}^{+0.0431}$ &  4.2722$_{-0.1262}^{+0.1250}$ &  1355.3707$_{-0.0602}^{+0.0601}$ &  0.5511$_{-0.0592}^{+0.0433}$ \\
  50270480 &            2 &   1.6514 $\pm$ 0.0034 &  1.7106 $\pm$ 0.1129 &        1354.5419 &  1.6514$_{-0.0057}^{+0.0056}$ &  1.8744$_{-0.2570}^{+0.2909}$ &  1354.5419$_{-0.0455}^{+0.0456}$ &  0.6734$_{-0.1947}^{+0.1126}$ \\
  80315892 &            2 &   6.1589 $\pm$ 0.0114 &  2.3433 $\pm$ 0.4739 &        1356.3507 &  6.1593$_{-0.0693}^{+0.0702}$ &  2.3539$_{-0.1868}^{+0.1900}$ &  1356.3505$_{-0.0523}^{+0.0520}$ &  0.6948$_{-0.0914}^{+0.0672}$ \\
 101991992 &            2 &   2.7375 $\pm$ 0.0032 &  1.6507 $\pm$ 0.1726 &         1355.861 &  2.7375$_{-0.0088}^{+0.0088}$ &  1.7981$_{-0.1439}^{+0.1319}$ &  1355.8611$_{-0.0359}^{+0.0358}$ &  0.6287$_{-0.1354}^{+0.0872}$ \\
 183231138 &            2 &   3.5578 $\pm$ 0.0036 &   2.558 $\pm$ 0.3320 &        1355.1669 &  3.5578$_{-0.0128}^{+0.0129}$ &   2.582$_{-0.1226}^{+0.1198}$ &  1355.1668$_{-0.0362}^{+0.0363}$ &  0.6665$_{-0.0645}^{+0.0520}$ \\
 184111843 &            2 &   3.0855 $\pm$ 0.0096 &  2.8029 $\pm$ 0.3234 &        1354.2508 &  3.0854$_{-0.0291}^{+0.0293}$ &  2.8583$_{-0.1121}^{+0.1117}$ &  1354.2509$_{-0.0444}^{+0.0439}$ &  0.3814$_{-0.1502}^{+0.0917}$ \\
 220570288 &            2 &   4.4427 $\pm$ 0.0057 &  1.8601 $\pm$ 0.3149 &         1356.232 &  4.4427$_{-0.0247}^{+0.0252}$ &  2.3565$_{-0.3356}^{+0.5918}$ &   1356.233$_{-0.0670}^{+0.0656}$ &  0.8089$_{-0.1072}^{+0.0902}$ \\
 281851658 &            2 &   6.1993 $\pm$ 0.0153 &  2.2804 $\pm$ 0.5746 &        1359.8757 &  6.2007$_{-0.0944}^{+0.0951}$ &  2.2441$_{-0.2588}^{+0.2911}$ &  1359.8751$_{-0.0685}^{+0.0689}$ &   0.632$_{-0.1798}^{+0.1245}$ \\
 408038524 &            3 &   1.2465 $\pm$ 0.0017 &  2.4323 $\pm$ 0.1706 &         1386.952 &  1.2465$_{-0.0021}^{+0.0022}$ &   4.565$_{-0.5071}^{+0.6234}$ &   1386.952$_{-0.0235}^{+0.0236}$ &  0.8776$_{-0.0613}^{+0.0894}$ \\
  10934226 &            4 &   4.8034 $\pm$ 0.0105 &  2.1807 $\pm$ 0.3692 &        1410.9917 &  4.8034$_{-0.0501}^{+0.0508}$ &  2.1443$_{-0.2090}^{+0.2219}$ &  1410.9919$_{-0.0517}^{+0.0516}$ &  0.5252$_{-0.2569}^{+0.1459}$ \\
  23138732 &            4 &   4.5577 $\pm$ 0.0057 &  3.3202 $\pm$ 0.5654 &        1410.9822 &  4.5579$_{-0.0263}^{+0.0260}$ &  3.3721$_{-0.1673}^{+0.1727}$ &   1410.982$_{-0.0376}^{+0.0380}$ &  0.6925$_{-0.0557}^{+0.0431}$ \\
  29960109 &            4 &   2.4917 $\pm$ 0.0077 &  2.2634 $\pm$ 0.2362 &        1411.6721 &  2.4918$_{-0.0192}^{+0.0192}$ &  2.2487$_{-0.1479}^{+0.1444}$ &  1411.6719$_{-0.0498}^{+0.0504}$ &   0.183$_{-0.1278}^{+0.1761}$ \\
  29960110 &            4 &   2.4918 $\pm$ 0.0075 &  2.1157 $\pm$ 0.2212 &         1411.673 &  2.4917$_{-0.0183}^{+0.0185}$ &  2.0664$_{-0.1802}^{+0.1790}$ &  1411.6728$_{-0.0479}^{+0.0485}$ &  0.2366$_{-0.1630}^{+0.2070}$ \\
  98796344 &            4 &   5.3591 $\pm$ 0.0106 &  1.3391 $\pm$ 0.2709 &        1412.7064 &  5.3596$_{-0.0565}^{+0.0558}$ &  1.2645$_{-0.1040}^{+0.1300}$ &  1412.7064$_{-0.0495}^{+0.0491}$ &  0.3311$_{-0.2228}^{+0.2354}$ \\
 161478895 &            4 &   4.8672 $\pm$ 0.0059 &  3.7196 $\pm$ 0.6281 &        1412.2682 &  4.8674$_{-0.0286}^{+0.0285}$ &  3.5509$_{-0.1196}^{+0.1247}$ &   1412.268$_{-0.0346}^{+0.0346}$ &  0.5627$_{-0.0443}^{+0.0485}$ \\
 206468250 &            4 &  5.5669 $\pm$ -0.2128 &  3.3042 $\pm$ 0.6695 &        1411.4297 &  5.5641$_{-0.0772}^{+0.0782}$ &  3.2233$_{-0.1408}^{+0.1481}$ &  1411.4356$_{-0.0439}^{+0.0445}$ &  0.6588$_{-0.0553}^{+0.0476}$ \\
 209457622 &            4 &   5.0354 $\pm$ 0.0079 &  1.6253 $\pm$ 0.3287 &        1414.9743 &  5.0353$_{-0.0395}^{+0.0400}$ &  1.5319$_{-0.2419}^{+0.2482}$ &  1414.9748$_{-0.0739}^{+0.0734}$ &  0.2612$_{-0.1798}^{+0.2482}$ \\
 259377017 &            4 &   5.6574 $\pm$ 0.0139 &  2.4869 $\pm$ 0.5032 &        1412.1471 &  5.6579$_{-0.0774}^{+0.0780}$ &  2.3529$_{-0.1629}^{+0.1838}$ &  1412.1475$_{-0.0587}^{+0.0590}$ &  0.3344$_{-0.2145}^{+0.1915}$ \\
 406478079 &            4 &     1.57 $\pm$ 0.0014 &  2.5025 $\pm$ 0.1749 &         1412.295 &    1.57$_{-0.0022}^{+0.0021}$ &  5.3054$_{-0.7614}^{+0.6264}$ &  1412.2949$_{-0.0221}^{+0.0222}$ &  1.0075$_{-0.0530}^{+0.0343}$ \\
   7688647 &            5 &   1.9359 $\pm$ 0.0035 &  2.4759 $\pm$ 0.2062 &        1439.1884 &  1.9359$_{-0.0067}^{+0.0066}$ &  2.7132$_{-0.0835}^{+0.0818}$ &  1439.1887$_{-0.0339}^{+0.0332}$ &   0.541$_{-0.0463}^{+0.0381}$ \\
  44669739 &            5 &   2.4111 $\pm$ 0.0048 &  2.7747 $\pm$ 0.2673 &        1438.0304 &   2.411$_{-0.0112}^{+0.0116}$ &  2.9053$_{-0.1014}^{+0.1039}$ &  1438.0305$_{-0.0372}^{+0.0373}$ &  0.6064$_{-0.0469}^{+0.0417}$ \\
 192833836 &            5 &   6.2378 $\pm$ 0.0193 &   3.487 $\pm$ 0.8803 &        1442.9738 &  6.2368$_{-0.1190}^{+0.1219}$ &  3.6226$_{-0.1609}^{+0.1534}$ &   1442.974$_{-0.0677}^{+0.0669}$ &  0.6221$_{-0.0663}^{+0.0505}$ \\
 220479565 &            5 &   3.6944 $\pm$ 0.0106 &  2.0914 $\pm$ 0.3052 &        1441.0207 &  3.6946$_{-0.0387}^{+0.0391}$ &  2.8577$_{-0.2144}^{+0.2376}$ &  1441.0206$_{-0.0546}^{+0.0549}$ &  0.7316$_{-0.0697}^{+0.0562}$ \\
 259377017 &            5 &   5.6605 $\pm$ 0.0157 &  2.5068 $\pm$ 0.5077 &        1440.4501 &  5.6604$_{-0.0880}^{+0.0893}$ &  2.2633$_{-0.1336}^{+0.1538}$ &  1440.4503$_{-0.0708}^{+0.0705}$ &  0.2968$_{-0.1941}^{+0.1794}$ \\
\bottomrule
\end{tabular}}
\end{table*}

\subsection{MCMC Transit Modeling}\label{sec:MCMC}
To model the transits of TCEs denoted as planet candidates following the vetting process described in \ref{sec:vetting}, we employ the use of the \textsf{exoplanet} Python package \citep{exoplanet:exoplanet}, and model various transits in each step of our Markov Chain Monte Carlo (MCMC) simulation. To model transits in general, \textsf{exoplanet} uses an analytical transit model, computed via the \textsf{STARRY} Python package \citep{exoplanet:luger18}. Our noise model contains the following free parameters: the orbital period P, time of mid-transit T0, planet-to-star radius ratio $\textrm R_{\textrm P} / \textrm R_{\textrm S}$ and impact parameter b. To define our prior distributions for the orbital period, we use a log-normal distribution with a mean value set to the TLS-detected period, and a standard deviation set to the TLS period error. For the prior distributions used for the mid-transit time, we use a normal distribution, with the mean value set to the TLS transit time, and the standard deviation set to the TLS transit duration. For our prior distribution of planet-to-star radius ratio, we use a uniform distribution ranging from 0.01 to $\textrm R_{\textrm P} / \textrm R_{\textrm S} + \sigma_{\textrm RP/RS}$. To calculate the uncertainty for the planet-to-star radius ratio, $\sigma_{\textrm RP/RS}$, we query the TIC to obtain the stellar radius and its uncertainty (R$_S$, $\sigma_{\textrm RS}$), and propagate the errors of the stellar radius, along with the planet radius uncertainty ($\sigma_{\textrm RP}$) as measured by TLS:

\begin{equation}\label{eq:rp_rs_err}
    \mathrm{\sigma_{\textrm RP/RS} = \frac{R_P}{R_S} \sqrt{
    \bigg( \frac{\sigma_{RP}}{R_P}\bigg)^2 +
    \bigg( \frac{\sigma_{RS}}{R_S}\bigg)^2}}
\end{equation}

To define our prior distributions for quadratic limbdarkening coefficients and the impact parameter, we utilize the distributions available within the \textsf{exoplanet} package. To determine the fiducial prior distribution for quadratic limb-darkening coefficients, we use a reparameterization of the two-parameter limb-darkening model to allow for efficient and uninformative sampling, as implemented by \citet{exoplanet:kipping13}. For the prior distributions of the impact parameter, we utilize a uniform distribution ranging from 0.01 to $ 1+ \textrm R_{\textrm P} / \textrm R_{\textrm S}$. Our adopted model parameter priors are listed in Table \ref{tbl:MCMC_priors}. For each transit model produced in the MCMC analysis, we oversample the light curve on a fine time grid, and numerically integrate over the exposure window to avoid smearing of the light curve due to TESS FFI’s 30-minute cadence.

\begin{figure}[htp]
    \includegraphics[width=\linewidth, trim=0cm 0.0cm 0cm 0.0cm, clip=true]{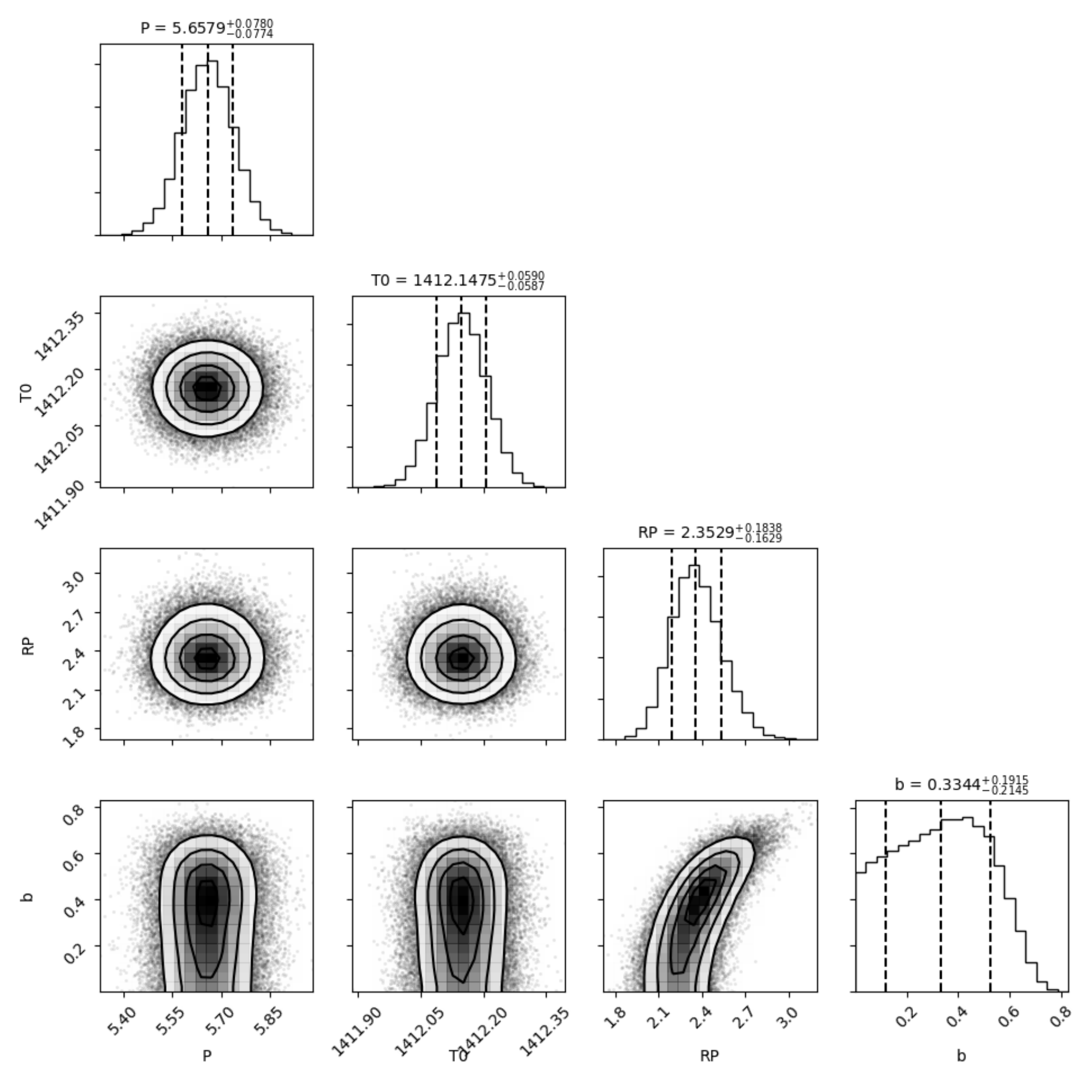}
    \caption{Distribution of transit model orbital parameter posteriors for planet TOI 270 c (TIC 259377017).}
    \label{fig:posterior_example}
\end{figure}

\begin{figure}[htp]
    \includegraphics[width=\linewidth, trim=0cm 0.0cm 0cm 0.0cm, clip=true]{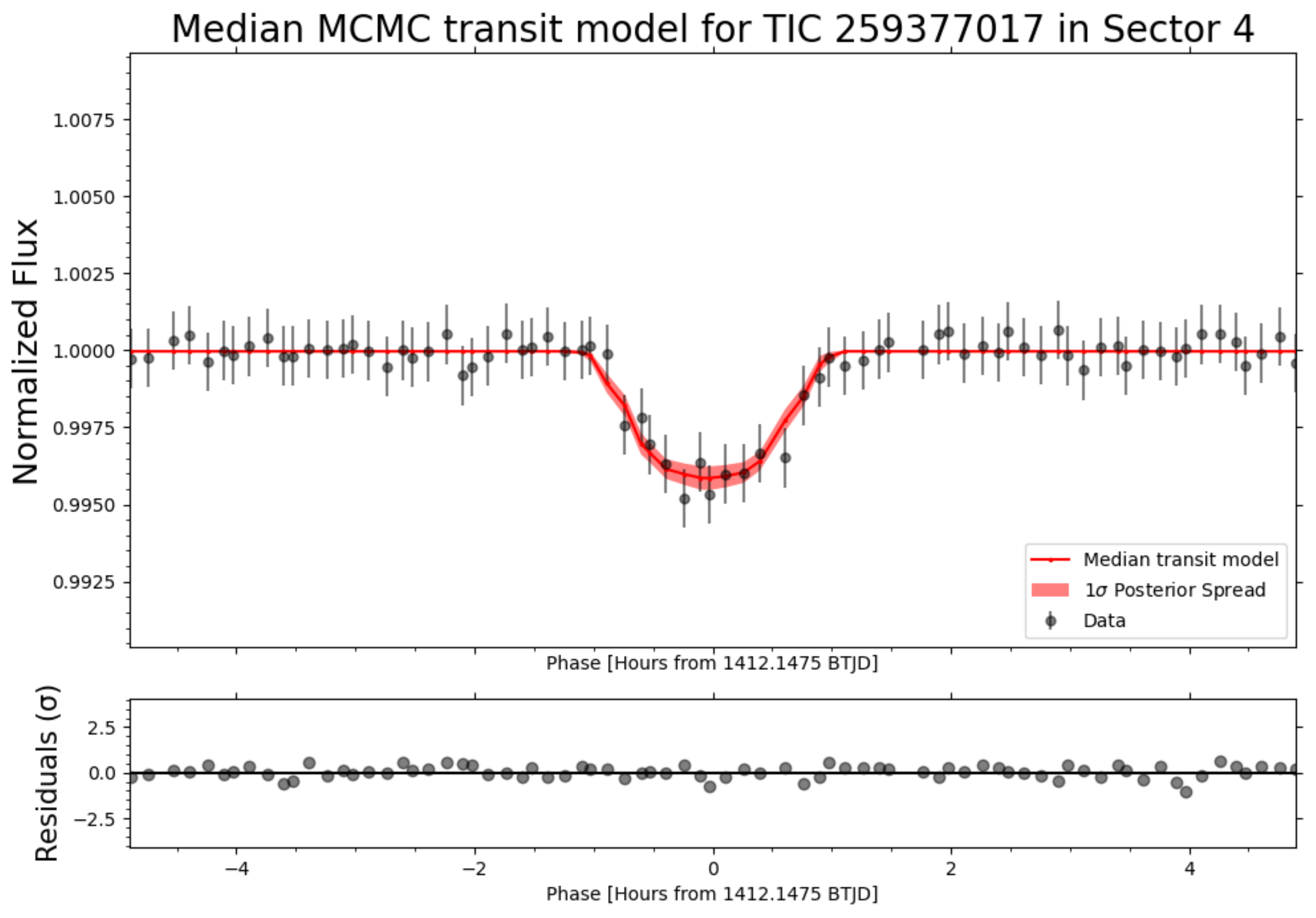}
    \caption{Median MCMC transit model for planet TOI 270 c (TIC 259377017). }
    \label{fig:mcmc_model}
\end{figure}

An initial maximum a posteriori (MAP) solution was found, and used to initialize the parameters sampled with an MCMC analysis. The MCMC sampling was performed using the No U-Turns step method \citep{Hoffman:2014}. We ran four chains with 10,000 tuning steps (tuning samples were discarded), and 12,500 draws with a target acceptance of 99\%, for a final chain length of 50,000 in each parameter. Once all our runs had been sampled, we then converted our posterior distributions of planet-to-star radius ratio to planet radii in Earth units. An example of the final posterior distributions for our free parameters, together with the median transit model from our MCMC analysis, can be seen in Figures \ref{fig:posterior_example} and \ref{fig:mcmc_model} for TOI 270 c (TIC 259377017).

\begin{table}[htp]
\begin{center}
\caption{Example of MCMC priors used for modeling planet candidates \label{tbl:MCMC_priors}}    
\begin{tabular}{c|c|cccc}
 Parameter &  Units &  Prior \\ \hline
 Period &  Days &     $\mathcal{N}$($\ln$ P , $\sigma_{\textrm P}$)\\ 
 T0 &  BTJD &    $\mathcal{N}$(T0 , Dur)\\ 
 $\frac{\textrm R_P}{\textrm R_S}$& --  & $\mathcal{U}$(0.01 , $\textrm R_P/R_S + \sigma_{\textrm RP}$ )  \\
 b & -- & $\mathcal{U}$(0.01 , $1+ R_P/R_S$)\\ 
\end{tabular}
\vspace{-0.1in}
\footnotetext[0]{\footnotesize{NOTE: For the input values of our prior distributions, we use the best-fit parameters determined by TLS: Period (P), Period error ($\sigma_P$), Transit Time (T$_0$), Transit Duration (Dur), Planet Radius (R$\mathrm{_P}$). We calculate the planet to star radius ratio ($\mathrm{R_P}/\mathrm{R_S}$ and its propagated error ($\sigma\mathrm{_{RP}}$) with Equation \ref{eq:rp_rs_err}.}}
\end{center}
\end{table}

\subsection{Transit Injection Analysis}\label{sec:TransInj}
\subsubsection{Transit Injection Recovery and Sensitivity}
\begin{figure*}[htp]
    \includegraphics[width=0.95\textwidth, trim=0cm 0.0cm 0cm 0.0cm, clip=true]{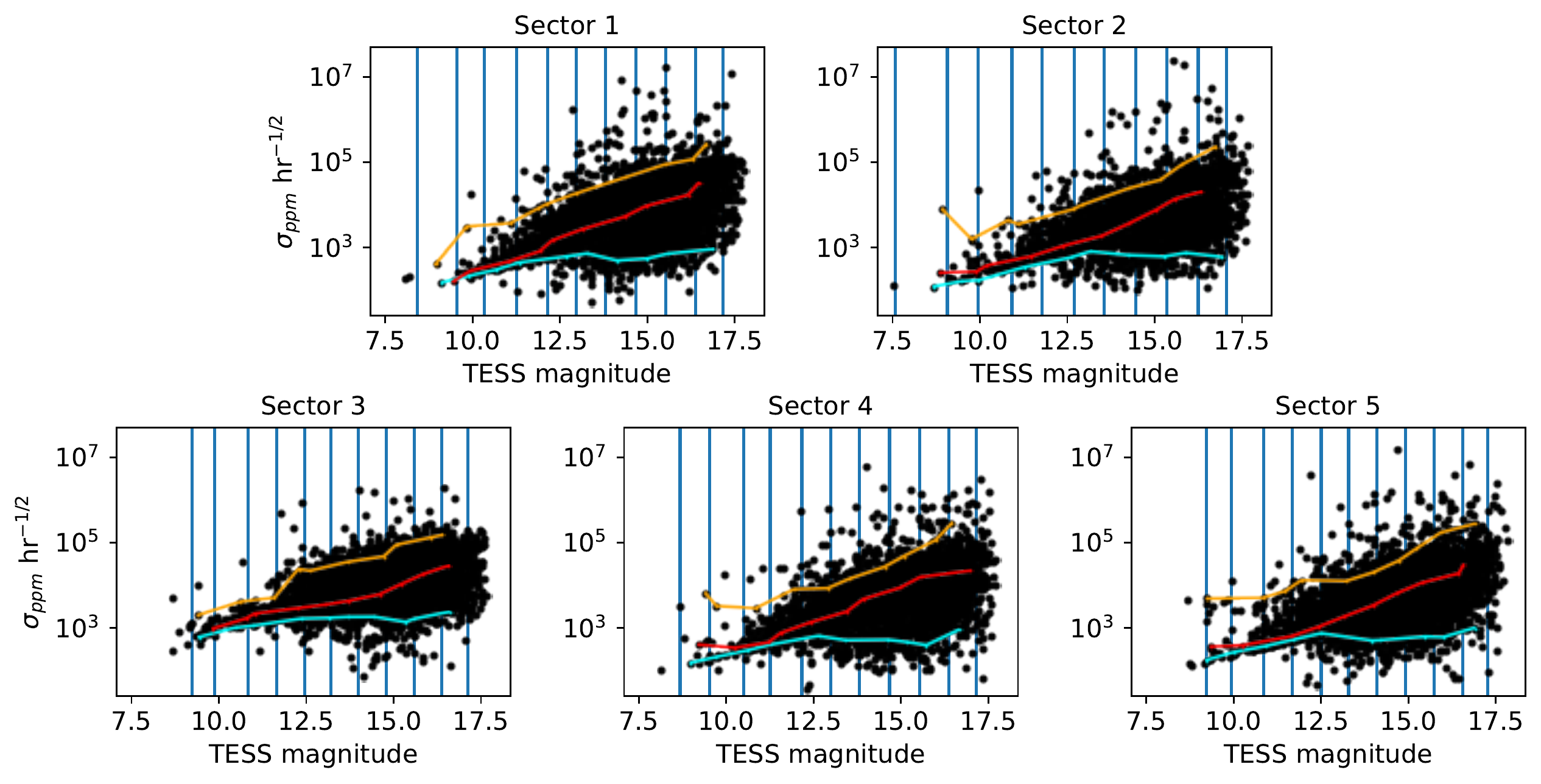}
    \caption{Combined differential photometric precision (CDPP) for the PLD corrected and smoothed light curves from sectors 1--5. To select stars representing the best, average, and worst quality of our light curves, we divided our data into bins of TESS magnitude, and selected those stars closest to the 0.025, 0.5, and 0.975 quantiles of their respective magnitude bins. Stars with CDPPs closest to the 0.025, 0.5, and 0.975 quantiles are indicated by the orange, red, and cyan line-connected points, respectively.}
    \label{fig:representative_stars}
\end{figure*}

To test the detection capability of our pipeline, we used a set of simulated data. In order to select stars that best represented the quality of our data, we looked at our CDPP noise metrics, and divided our light curves into 11 bins of TESS magnitudes, as shown in Figure \ref{fig:representative_stars}. For each magnitude bin, we then identified light curves with CDPPs closest to the 0.025, 0.5, and 0.975 quantiles as representations of the best, average, and worst quality data, indicated by cyan, red, and orange lines in Figure \ref{fig:representative_stars}, respectively. For each sector, we selected 30 stars to represent our sample. We randomly sampled uniform distributions of orbital periods ranging from 0.5 to 9 days, and planet radii from 0.5 to 11 Earth radii, and calculated simulated transit models using the \textsf{BATMAN} Python package \citep{Kreidberg:2015}.
 
\begin{figure*}[htp]
    \includegraphics[width = 0.95\textwidth,scale=0.25, trim=0cm 0.0cm 0cm 0.0cm, clip=true]{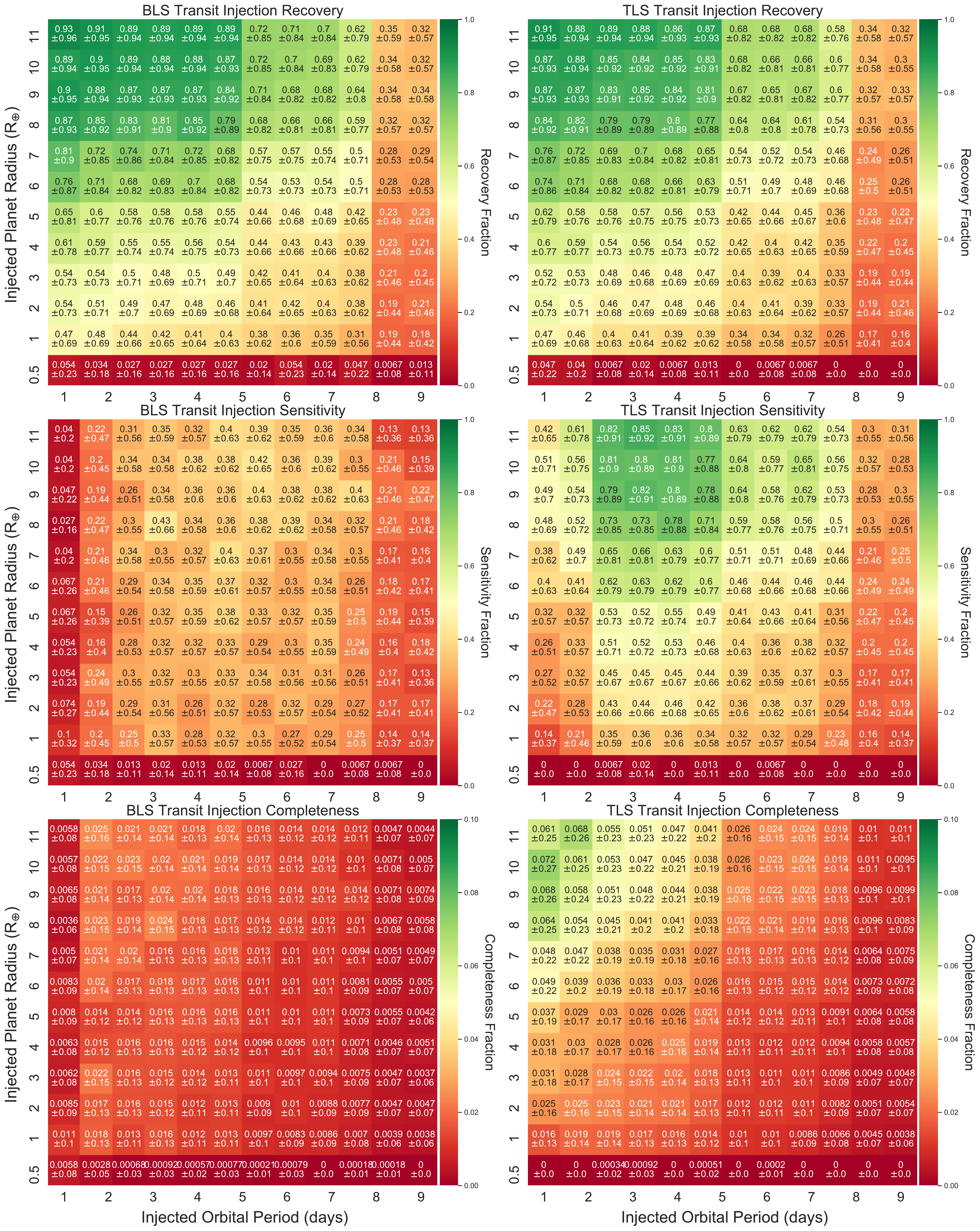}
    \caption{Transit injection analysis of 21,600 injected light curves across TESS sectors 1 -- 5 as described in Section \ref{sec:TransInj}. Top row: fraction of detected injected transits from both BLS and TLS that meet our recovery criteria, where the recovered period is within 1\% of injected period. Middle row: fraction of detected injected transits from both BLS and TLS that meet our sensitivity criteria, where the recovered period is within 1\% of injected period, and injected radius is within recovered planet radius $\pm$ planet radius uncertainty. Bottom row: BLS and TLS completeness maps, representing the results of the product of detection sensitivity and geometric transit probability, as described in Section \ref{sec:completeness_occurrence}. The uncertainty of these 2D maps are Poisson counting errors for each cell of orbital period and planet radius.}
    \label{fig:transit_injection}
\end{figure*}

Once aperture photometry has been performed on the FFI cutout of a representative star, we inject our simulated transits, starting randomly within the first orbit of TESS in a given sector, and then applying the rest of our pipeline’s processes, as described in Section \ref{sec:data_reduction}. For each of the 30 representative stars, we used 12 randomly selected orbital periods and planet radii, for a total of 21,600 transit injections across TESS sectors 1--5. Our criteria for a successfully recovered transit is based on the orbital period of the transits detected by TLS being within 1\% of the injected period. To test the sensitivity of our pipeline, we add an extra criteria to determine whether the injected planet radius falls within the measured error of the TLS-modeled planet radius. We display our 2D injection recovery rate and detection sensitivity maps in Figure \ref{fig:transit_injection}. Based on our transit injection analysis, we can see that in the 1--9 day regime operated by our transit search, our pipeline is sufficiently sensitive to detect more than 30\% of transiting planets with radius $>$1 \RE, and with periods between two and five days. For periods between two and eight days, we maintain a sensitivity of $>$30\%, approximately, for planet radii $>$4 \RE. For injected transits across the full period space explored, and radii $>$1 \RE, the lower limit of our detection sensitivity is $\sim$14\%.
 
\subsubsection{Survey Completeness and Occurrence Rates for M-dwarf stars}\label{sec:completeness_occurrence}
The derivation of exoplanet occurrence rates requires the distribution of planet detections to be corrected for imperfect survey completeness. To account for detectable nontransiting planets, we compute the geometric transit probability for each star (\textit{n}) at each orbital period, and planet radius grid cell (\textit{i,j}) in our 2D sensitivity maps as

\begin{equation}
    \mathrm{P_{transit}(n,i,j) = \frac{ R_{star,n} + R_{Planet,j}}{(\frac{ \mathrm{G~M_{star,n}~P_{i}^2}}{4\pi^2})^{\frac{1}{3}}}}
\end{equation}

The product of our 2D detection sensitivity maps with
geometric transit probability yields a 2D completeness map, which can be seen in Figure \ref{fig:transit_injection}. To estimate the occurrence rate of planets per star for our M-dwarf host stars, we choose to use the TLS 2D completeness map, which is more complete, due to its higher detection sensitivity. For the period–radius parameter space explored in our survey, and the five TESS sectors searched, there are only a few dozen planet candidates, occupying a small region of the period–radius parameter space. Without a more diverse catalog of planet candidates occupying more of the period–radius parameter space, creating a meaningful 2D occurrence rate map is difficult. For this reason, we instead use an integrated occurrence rate to estimate, on average, the number of planets per star for M-dwarf host stars.

By counting the number of times our planet candidates occur within our period and planet radius grid cells, N$_{\text{detected}}$(P,R$_{\text{P}}$), we can calculate the integrated occurrence rate, O(P,R$_{\text{P}}$), as a double integral over the period and planet radius space:

\begin{equation}
    \mathrm{O(P,R_{P})=\int \int  \frac{N_{detected}(P,R_{P})}{N_{stars} \times Completeness(P,R_{P})} ~ dPdR_{P}}
\end{equation}

For the full parameter space of P $\in$ [1, 9] days and R$_{\text{P}}$ $\in$ [0.5, 11]~\RE, we calculate the integrated occurrence rates as 1.61 $\pm$ 1.27 planets per star. In the same range of periods, but with R$_{\text{P}}$ $\in$ [0.5, 4]~\RE, we calculate the integrated occurrence rates as 1.45 $\pm$ 1.2 planets per star. If we also include all other confirmed planets, TOI and DIA planet candidates in the range P $\in$ [1, 9] days and R$_{\text{P}}$ $\in$ [0.5, 11]~\RE, we calculate the integrated occurrence rates as 2.49 $\pm$ 1.58 planets per star. In the same range of periods, but with R$_{\text{P}}$ $\in$ [0.5, 4]~\RE, we calculate the integrated occurrence rates as 3.62 $\pm$ 1.9 planets per star. Previous studies exploring occurrence rates for M-dwarf host stars observed in the Kepler and K2 missions (\citealt{Morton:2014}, \citealt{Dressing:2015}, \citealt{Gaidos:2016}, \citealt{Hardegree_Ullman:2020}, \citealt{Cloutier:2020} and \citealt{Hsu:2020}) found occurrence rates ranging in value from $\sim$ 1--22.5 planets per star for periods $<$200 days, which is in agreement with our results. A more detailed comparison can be found in Section \ref{sec:yield_and_occurrence_rate_comparison}.

\section{Discussion}\label{sec:discussion}

\subsection{Prospects for Mass Characterization via the Radial Velocity Method}
One of the primary goals of the TESS mission is to discover at least 50 planets with radii $<$ 4 \RE, and with measured masses via radial velocity follow-up observations. Many of our planet candidates have radii below 4 \RE~(See Figure \ref{fig:pc_demographics} and Table \ref{tbl:MCMC_results}). In an effort to infer the expected radial velocity semi-amplitude of our planet candidates, we utilize the empirical mass–radius relation from \citet{Chen:2016} to compute the planetary mass of our planet candidates. We assume circular orbits (\textit{e = 0, i = 90\degrees}) to calculate the expected radial velocity semi-amplitude as

\begin{equation}
    \mathrm{K = \bigg( \frac{2\pi G}{Period} \bigg)^{1/3} \bigg( \frac{M_{Planet}\sin{i}}{{M_{Star}}^{2/3}} \bigg)
    \bigg(\frac{1}{\sqrt{1-e^2}}\bigg)}
\end{equation}

The expected radial velocity semi-amplitude values are
displayed in Table \ref{tbl:followup_ranking} and Figure \ref{fig:expected_rvs}. Some radial velocity
spectrometers, such as ESPRESSO, NEID, and EXPRES, are
able to remain stable at the level of a few tens of cm s$^-1$. However, due to photon noise and stellar activity, radial velocities can be limited to maintaining an accuracy of a few m s$^-1$ \citep{Fischer:2016}. All of our planet candidates have expected radial velocities ranging from 1.8 to 43.9 m s$^-1$, most of which are above this sensitivity limit.

\subsection{Prospects for Follow-Up Atmospheric Characterization}
To further assist in providing an assessment of suitability for follow-up atmospheric characterization, we also calculate the transmission spectroscopy metric (TSM) and emission spectroscopy metric (ESM), as outlined in Equations (1) and (4) of  \citealt{Kempton:2018}. The TSM is proportional to the expected transmission spectroscopy S/N, and, together with the inferred planetary mass data, is calculated as

\begin{equation}
    \mathrm{TSM  = C \times \frac{R_{planet}^3 T_{eq,planet}}{M_{planet} R_{star}^2} \times 10^{-J~mag / 5}}
\end{equation}

where \textit{C} is defined as:

\begin{equation}
    C =
    \begin{cases}
      0.19, & \text{if}\ \mathrm{R_{planet}}<1.5~\mathrm{\RE} \\
      1.26, & \text{if}\ 1.5~\mathrm{\RE}<\mathrm{R_{planet}}<2.75~\mathrm{\RE} \\
      1.28, & \text{if}\ 2.75~\mathrm{\RE}<\mathrm{R_{planet}}<4~\mathrm{\RE} \\
      1.15, & \text{if}\ 4~\mathrm{\RE}<\mathrm{R_{planet}}<10~\mathrm{\RE} \\
    \end{cases}
  \end{equation}

ESM is proportional to the expected S/N of a James Webb Space Telescope secondary eclipse detection at mid-infrared wavelengths, and is calculated as

\begin{equation}
\small
    \mathrm{ESM = 4.29\times10^{6} \times \frac{B_{7.5(T_{day})}}{B_{7.5(T_{eff,star})}} \times \frac{R_{planet}}{R_{star}}^{2} \times 10^{-K~mag / 5}}
\end{equation}

where \textit{B} is Planck’s function, evaluated for a given temperature, at a representative wavelength of 7.5 $\mu$m and \textit{T$\mathrm{_{day}}$} is the dayside temperature in Kelvin which we calculate as 1.1 times the equilibrium temperature of the planet (\textit{T$\mathrm{_{eq,planet}}$}). Assuming albedo (A=0) and heat redistribution factor (f=1), we calculate T$\mathrm{_{eq,planet}}$ as

\begin{equation}
    \mathrm{T_{eq,~planet}}  = \mathrm{T_{eff}} \sqrt{\frac{\mathrm{R_{star}}}{a}} \bigg( \frac{1-A}{4f} \bigg)^{1/4} 
\end{equation}

Moreover, we also calculate the S/N of our detected transit events as a function of transit depth, photometric precision, number of transits, transit duration, and cadence:

\begin{equation}\label{eq:SNR}
    \mathrm{SNR  = \frac{\delta_{transit}}{\sigma_{CDPP}}\sqrt{\frac{N_{transits}~T_{duration}}{cadence}}}
\end{equation}

\begin{figure}[tp!]
    \includegraphics[width=\columnwidth]{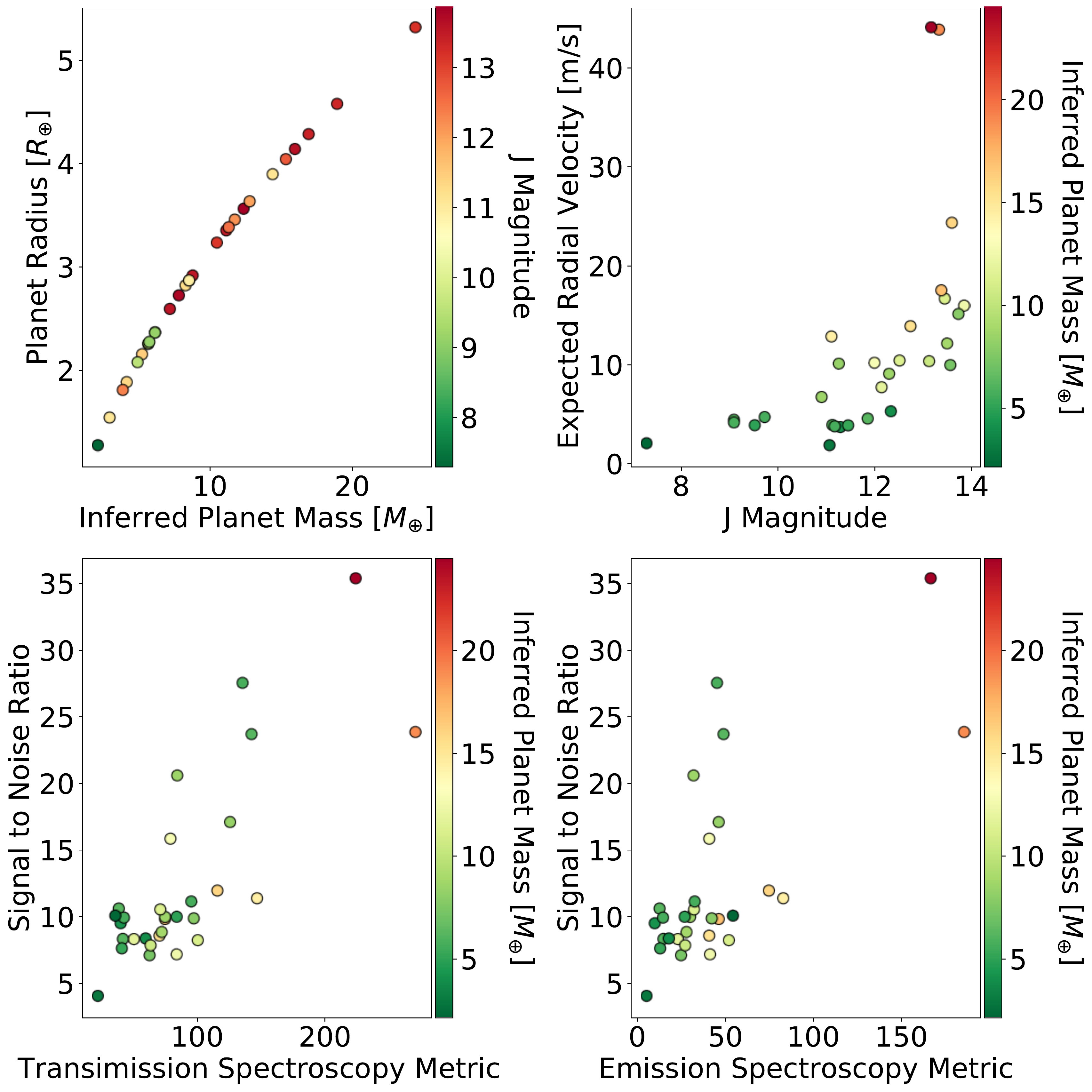}
    \caption{Upper left: using the planet mass–radius relations from \citealt{Chen:2016}, we calculated the inferred planetary mass of our planet candidates. Upper right: expected values of the radial velocity semi-amplitudes for our planet candidates as a function of J-band magnitude. Lower left and right: based on the results of inferred planet mass, we calculate the transmission and emission spectroscopy metrics introduced by \citealt{Kempton:2018} as functions of the S/N.}
    \label{fig:expected_rvs}
\end{figure}

The TSM, ESM, and S/N values are displayed in Table \ref{tbl:followup_ranking} and Figure \ref{fig:expected_rvs}. For planet candidates with S/Ns $>$7, 23 of our new candidates have TSMs $>$38, and ESMs $>$10, making them promising targets for follow-up characterization.

\begin{table*}[htp!]
\centering
\caption{ATMOSPHERIC CHARACTERIZATION RANKINGS FOR FOLLOW-UP OBSERVATIONS}
\label{tbl:followup_ranking}
\resizebox{0.95\textwidth}{!}{\begin{tabular}{rrrrrrrrrrrrr}
\toprule
    TIC ID &  TESS Sector &    RA [deg] &  Dec [deg] &  TESS magnitude &  J magnitude &  K magnitude &  R$_{P}$ [R$_{\oplus}$] &  MP [M$_{\oplus}$] &        SNR &  Doppler Semi-Amplitude [m/s] &         TSM &         ESM \\
\midrule
 139456051 &            1 &  336.783859 & -47.229283 &        15.34970 &       13.600 &       12.714 &           4.128602 &          15.986691 &  11.874213 &                     24.230491 &  116.030822 &   74.903532 \\
 143996634 &            1 &  335.753765 & -53.327428 &        14.43140 &       12.745 &       11.965 &           4.030119 &          15.347696 &   8.495298 &                     13.790703 &   70.734930 &   41.019245 \\
 234299676 &            1 &  358.611649 & -61.818088 &        13.65490 &       12.148 &       11.294 &           3.445461 &          11.758089 &   8.235459 &                      7.627944 &   50.721908 &   23.240705 \\
 281461138 &            1 &    5.988896 & -55.528776 &        13.01530 &       11.112 &       10.245 &           3.884640 &          14.417693 &  11.298390 &                     12.750291 &  147.036441 &   83.062000 \\
 290048573 &            1 &  319.188844 & -25.786300 &        12.67990 &       11.265 &       10.514 &           2.810214 &           8.301153 &  17.013639 &                     10.010448 &  125.998483 &   46.421523 \\
 389662862 &            1 &  323.275188 & -42.574423 &        15.24620 &       13.452 &       12.529 &           3.341961 &          11.163913 &   8.154024 &                     16.573354 &  100.699639 &   52.192536 \\
   9632592 &            2 &  357.198331 &  -7.863519 &        15.01790 &       13.382 &       12.562 &           4.272193 &          16.950659 &   9.739416 &                     17.412758 &   75.026066 &   46.449747 \\
  50270480 &            2 &   29.168067 & -71.953540 &        12.46420 &       11.296 &       10.550 &           1.874404 &           4.167672 &   9.421817 &                      3.608885 &   40.362280 &   10.097319 \\
  80315892 &            2 &    8.196481 & -44.715797 &        13.38990 &       11.862 &       11.020 &           2.353903 &           6.152647 &   8.246789 &                      4.467794 &   42.069438 &   15.043026 \\
 101991992 &            2 &   15.369636 & -46.542464 &        14.01590 &       12.340 &       11.565 &           1.798130 &           3.891827 &   8.286213 &                      5.196333 &   59.920862 &   18.107811 \\
 183231138 &            2 &  357.055156 & -38.684239 &        15.28160 &       13.568 &       12.769 &           2.581981 &           7.200179 &   7.023134 &                      9.866667 &   63.008718 &   25.058121 \\
 184111843 &            2 &  359.132827 & -36.364779 &        13.82280 &       12.304 &       11.477 &           2.858288 &           8.559577 &   9.896756 &                      8.988186 &   75.106971 &   30.192509 \\
 220570288 &            2 &   46.572050 & -64.829543 &        12.45500 &       11.131 &       10.291 &           2.356532 &           6.160028 &  10.528961 &                      3.820460 &   38.934106 &   12.855060 \\
 281851658 &            2 &   14.821597 & -59.204671 &        12.69350 &       11.180 &       10.328 &           2.244115 &           5.672862 &   9.847903 &                      3.685254 &   42.977857 &   14.839293 \\
 408038524 &            3 &   16.778423 &  -6.879304 &        15.05720 &       13.334 &       12.480 &           4.565040 &          18.945825 &  23.763564 &                     43.716183 &  270.929945 &  185.734208 \\
  10934226 &            4 &   42.793311 &  -1.337914 &        12.87200 &       11.458 &       10.657 &           2.144324 &           5.250814 &   7.541693 &                      3.779542 &   41.305156 &   13.173706 \\
  23138732 &            4 &   46.255879 & -11.093959 &        14.06900 &       12.519 &       11.702 &           3.372111 &          11.335044 &  10.461356 &                     10.325772 &   71.365712 &   32.434492 \\
  29960109 &            4 &   42.249290 & -14.537842 &        11.20030 &        9.733 &        8.875 &           2.248694 &           5.693698 &  11.058413 &                      4.619607 &   95.623689 &   32.799796 \\
  29960110 &            4 &   42.246970 & -14.537484 &        10.94730 &        9.528 &        8.646 &           2.066381 &           4.931000 &   9.909295 &                      3.792616 &   84.208728 &   27.087291 \\
  98796344 &            4 &   45.464137 & -16.593340 &         8.84294 &        7.294 &        6.496 &           1.264512 &           2.139913 &   9.992555 &                      1.973426 &   36.164831 &   54.524599 \\
 161478895 &            4 &   73.302476 & -44.393777 &        15.61980 &       13.853 &       13.009 &           3.550880 &          12.378486 &   7.091324 &                     15.862767 &   84.175164 &   41.503154 \\
 206468250 &            4 &   57.037751 & -50.795835 &        14.67190 &       13.130 &       12.341 &           3.223327 &          10.499727 &   7.767496 &                     10.257237 &   63.799177 &   27.414030 \\
 209457622 &            4 &   46.612506 & -33.942109 &        12.37050 &       11.073 &       10.235 &           1.531888 &           2.964851 &   3.977976 &                      1.778832 &   22.528110 &    5.474758 \\
 259377017 &            4 &   68.415501 & -51.956232 &        10.49810 &        9.099 &        8.251 &           2.352926 &           6.149659 &  23.604753 &                      4.343076 &  142.742900 &   49.139620 \\
 406478079 &            4 &   53.433858 & -22.214483 &        14.95450 &       13.172 &       12.359 &           5.305437 &          24.450171 &  35.294599 &                     43.955433 &  224.217887 &  166.763819 \\
   7688647 &            5 &   65.362813 & -39.013727 &        15.57150 &       13.735 &       12.922 &           2.713195 &           7.832774 &   9.791784 &                     15.018084 &   97.561991 &   42.452327 \\
  44669739 &            5 &   59.891484 & -28.249734 &        14.91290 &       13.502 &       12.702 &           2.905316 &           8.797528 &   8.766469 &                     12.044053 &   72.519333 &   28.164262 \\
 192833836 &            5 &   83.103441 & -39.833811 &        13.69400 &       12.001 &       11.096 &           3.622602 &          12.805514 &  15.762378 &                     10.096336 &   79.303364 &   41.048627 \\
 220479565 &            5 &   75.846088 & -54.177201 &        12.29580 &       10.909 &       10.100 &           2.857674 &           8.551350 &  20.506703 &                      6.641619 &   84.642565 &   32.150844 \\
 259377017 &            5 &   68.415501 & -51.956232 &        10.49810 &        9.099 &        8.251 &           2.263309 &           5.756908 &  27.452164 &                      4.065023 &  135.695670 &   45.460875 \\
\bottomrule
\end{tabular}}
\end{table*}

\subsection{Comparison to Exoplanet Yield Estimates and Occurrence Rates}\label{sec:yield_and_occurrence_rate_comparison}
Previous studies by \citealt{Sullivan:2015}, \citealt{Barclay:2018} and \citealt{Ballard:2019} have predicted that the TESS mission will discover thousands of planets over the course of its two-year mission. In terms of M-dwarf stars, \citealt{Sullivan:2015} predict the detection of roughly 1,700 transiting planets from about 200,000 stars observed in a two-minute cadence, with about 419 planets with radii $>~2$~\RE~orbiting M-dwarfs. \citealt{Barclay:2018} predict that TESS will find 496 planets orbiting M-dwarfs, of which 371 orbit stars observed in a two-minute cadence. \citealt{Ballard:2019} performed yield estimations focused on M-dwarfs with effective temperatures ranging from 3200--4000 K, and postulated that TESS would find approximately 1274 planets, orbiting 1026 M-dwarf stars.

In this study, we found 33,054 nearby M-dwarf stars meeting our selection criteria, as outlined in Section \ref{sec:targets}. Applying selection cuts of $2300~K<\mathrm{T_{eff}}<4000~K$, $\mathrm{R_{Star}}<0.5~\RS$,  $\mathrm{M_{Star}}<0.5 \MS$ and  $\mathrm{logg}>3$ to the TESS Candidate Target List\footnote{https://filtergraph.com/tess\_ctl} (CTL Version 8), we find 138,962 M-dwarf stars within 100 pc, and 1,562,038 M-dwarf stars in total. Using the predicted number of M-dwarf TESS planets from \citealt{Ballard:2019}, the expected planet yield from sectors 1 -- 5 is 1,274 planets $\times$ 33,054 stars / 1,562,038 stars $\sim$ 27 $\pm$ 5 planets where the uncertainty is $\mathrm{\sqrt{N}}$. Our yield of 29 planet candidates in TESS sectors 1–5 matches very well with this predicted yield estimate. Propagating our rates of detection to the M-dwarf stars in the CTL, we expect to detect (29 planets /  33,054 stars) $\times$ 138,962 stars $\sim$ 122 $\pm$ 11 planets of nearby stars from FFI data over the two-year TESS mission. If we extend our search to more distant stars ($>$100 pc), and assume similar rates of detection, we can expect to detect (29 planets /  33,054 stars) $\times$ 1,562,038 stars $\sim$ 1370 $\pm$ 37 planets total from the FFI data for the two-year TESS mission
 
Various studies over the past few years have focused on occurrence rates for M-dwarfs, using confirmed planets and planet candidates from the NASA Kepler and K2 missions. \citealt{Morton:2014} found an occurrence rate of Kepler M-dwarf planet hosts to be 2 $\pm$ 0.45 planets per star for P $< 150$ days and R$_{\text{P}} \in [0.5, 4]$~\RE. \citealt{Gaidos:2016} found an occurrence rate of 2.2 $\pm$ 0.3 planets per star for P $\in [0.5,180]$ days and R$_{\text{P}} \in [1, 4]$~\RE. \citealt{Dressing:2015} found Kepler M-dwarf planet hosts to have an occurrence rate of 2.5 $\pm$ 0.2 planets per star for P~$<200$ days and R$_{\text{P}} \in [1, 4]$~\RE; for Earth-sized planets (R$_{\text{P}} \in [1, 1.5]$~\RE) and Super-Earths (R$_{\text{P}} \in [1.5, 2]$~\RE) in the ranges P $<50$ days, the occurrence rate is 0.56 $\pm$ 0.06 and 0.46$^{+0.07}_{-0.05}$ planets per star, respectively. \citealt{Hardegree_Ullman:2020} explored occurrence rates of Kepler M-dwarfs by spectral type and computed an occurrence rate of 1.19$^{+0.7}_{-0.49}$ planets per star in the parameter space of P $\in [0.5,10]$ days and R$_{\text{P}} \in [0.5, 2.5]$~\RE. \citealt{Cloutier:2020} explored the parameter space in the ranges of P $\in [0.5,100]$ days and R$_{\text{P}} \in [0.5, 4]$~\RE and found occurrence rates for the Kepler and K2 missions in the ranges of 2.485$\pm$0.32 and 2.26$\pm$0.38 planets per star, respectively. \citealt{Hsu:2020} found occurrence rates of Kepler M-dwarf planet hosts to be  4.2$\pm0.6$ or 8.2$^{+1.2}_{-1.1}$, depending on their choice of priors in sampling their distribution, for the ranges P $\in [0.5,256]$ days and R$_{\text{P}} \in [0.5, 4]$~\RE. \citealt{Hsu:2020} also found that for small planets at short periods in the range P $\in [2,32]$ days and R$_{\text{P}} \in [1, 2.5]$~\RE, their occurrence rate is 0.9$^{+0.2}_{-0.1}$ or 1.6$^{+0.3}_{-0.2}$ (depending on prior chosen).

As mentioned in Section \ref{sec:completeness_occurrence}, our integrated occurrence rate for the ranges P $\in [1,9]$ days and R$_{\text{P}} \in [0.5, 11]$~\RE~is 2.49 $\mathrm{\pm 1.58}$ planets per star, when including known transiting planets, TOI, and DIAmante planet candidates. This occurrence rate is within the uncertainty of the estimations of occurrence rates from previous studies in similar period–radius parameter space. Of our 29 planet candidates, 24 of them have planet radii $<4$ \RE with measurable planet masses (see Table \ref{tbl:followup_ranking}), and can
thus contribute to fulfilling the level one science requirement of the TESS mission\footnote{https://tess.mit.edu/followup/}.

\subsection{Limitations of this Survey Strategy}\label{sec:Limitations}
Although there are more M-dwarfs observed by TESS with 30-minute cadences than with two-minute cadences, the lack of time resolution limits the types of transit that can be detected. Using Equations (\ref{eq:T_dur}) and (\ref{eq:SNR}), for a planet with a radius of 1 \RE~transiting a small star with $\mathrm{M = 0.1~\MS}$, $\mathrm{R = 0.1~\RS}$ once per day, the transit duration is about 17 minutes with a SNR $\sim$ 127 for a two-minute cadence and a SNR $\sim$ 33 for a 30-minute cadence. For a larger star ($\mathrm{M = 0.5~\MS}$, $\mathrm{R = 0.5~\RS}$) and the same transiting planet, the transit duration is about 47 minutes with a SNR about 8 for a two-minute cadence and a SNR about 2 for a 30-minute cadence. 

With sufficient coverage of observed transit events, it is possible to detect these short-duration, small-transit events in FFI data, but with the various instrumental systematics and astrophysical noise presented by each of these faint targets, these 1–2 data points per transit may be removed during data processing, owing to common data-reduction practices such as smoothing or outlier removal. With our single-sector approach to searching for transit events, the types of transits we can detect with the 30-minute cadence data is somewhat limited, as shown in Figure \ref{fig:transit_injection} where we maintain a detection sensitivity of roughly 30\% for radii larger than 1 \RE. In years three and four of the TESS extended mission, the FFIs will be observed with a 10-minute cadence, which will certainly increase the detection of these types of transit events, as they are likely to have higher S/Ns due to the improved time resolution.

Another challenge encountered in this survey lies in ensuring that our automatically selected apertures remained on-target throughout the observations in each sector of data. As mentioned in Section \ref{sec:SAP_photometry}, we utilized a bivariate quadratic function to approximate the core of a point-spread function for each FFI cutout, where we chose pixels that were brighter than 7.5 standard deviations above the median brightness of the time series. For isolated targets, this works fairly well, but for crowded fields, particularly for dim targets (TESS magnitude $>$15), our centroiding procedure often fails, and the selected aperture centers on other bright stars in the 11 $\times$ 11 pixel cutouts. 

\subsection{Future Work}

Early in the testing phase of our pipeline’s application, we noticed that each sector of TESS data has varying degrees of instrumental effects due to telescope jitter or glare in the TESS field of view from Earth and/or the Moon near the beginning or ends of data collection in each orbit of the satellite. For this reason, we opted to focus on performing transit searches on individual sectors, thereby restricting the upper limit in terms of the longest periods we could search per sector, to  about nine days. This choice was cautiously selected early in the development of our pipeline, so as to avoid increasing the amount of false positives encountered via a multi-sector approach that could possibly be attributed to residual instrumental effects in our processed light curves.

We intend to continue our survey to cover TESS sectors 6--13 in the southern hemisphere, and 14--26 in the northern sectors, in order to gain a longer baseline of coverage for our M-dwarf target lists, and to extend the range of periods to search for transits using a multi-sector approach. With a longer baseline of observations, we can also consider searching for multiple planets per target with a higher degree of confidence than in our single-sector approach, by analyzing other peaks in the TLS power spectrum with strong SDEs (see Appendix \ref{sec:appendix_comparison}, TOI 270). We also intend to vet those threshold crossing events with lower SDEs in the range of 6--10, in order to potentially find transit detections that may have been missed due to the use of a higher threshold. 

Although we were cautious in our promotion of planet candidates from TCEs, there is still an element of human judgement, and the potential for bias. As a quick check, and an alternative analysis to our vetting report, we used the automated EDI-Vetter Unplugged tool, which uses our TLS outputs to conduct false-positive tests, as described in Section \ref{sec:vetting}. Additional community tools, such as TESS-ExoClass (TEC\footnote{https://github.com/christopherburke/TESS-ExoClass}) or Discovery and Vetting of Exoplanets (DAVE, \citealt{Kostov:2019b}b) are vetting tools built upon the Kepler vetting algorithm RoboVetter \citep{Coughlin:2016}, and were developed for use in K2 \citep{Hedges:2019} and TESS data (\citealt{Crossfield:2019}, \citealt{Kostov:2019a}a). For our future endeavors, to minimize opportunities for human selection bias, and to reduce the overall human vetting time, we may explore automating our vetting process, or the use of tools such as DAVE or TEC for the first round of vetting in our overall vetting process.

In future work, we may also revisit our centroiding process, or employ a multi-aperture approach similar to that used in other FFI pipelines (\citealt{Feinstein:2019}; \citealt{Nardiello:2019}; \citealt{Bouma:2019}; \citealt{Montalto:2020}). We currently keep track of the change in pixel directions during any detected transit-like events, and exclude threshold crossing events with change in pixel direction of more than five standard deviations, to help reduce the amount of false-positive detections, but an alternative approach to centroiding may prove more consistent for targets fainter than TESS magnitude $>$15.

All of these candidates were observed by TESS roughly two years ago, and the southern hemisphere is already being observed once again by TESS in year three of its extended mission. In the extended TESS mission, FFIs will have an improved time resolution of 10-minute cadences, whereas years one and two, the time resolution was 30 minutes. After working through the data from years one and two, we also intend to extract light curves from the year three data, in order to revisit our planet candidates based on an improved time resolution. Armed with a longer baseline, by virtue of analyzing the FFI data from both the TESS mission and its extended mission, we also intend to revisit the detection recoverability, sensitivity, and survey completeness of our \textsf{NEMESIS} pipeline for longer orbital periods. With additional 
future planet candidate detections, we can also revisit our estimation of planet occurrence rates for M-dwarf host stars with improved resolution in the orbital period–planet radius domain.

\section{Conclusion}\label{sec:conclusion}
In this work, we have developed a new pipeline called \textsf{NEMESIS} to extract photometry from the TESS FFIs, remove instrumental systematics and long-term stellar variations, and conduct transit searches. We applied the \textsf{NEMESIS} pipeline to
33,054 M-dwarfs located within 100 parsecs, and observed by TESS in sectors 1--5. We then presented our transit search approach, and discussed the overall transit-detection recoverability, sensitivity, and completeness of our pipeline via a transit injection analysis. Our methods have proved successful, with the detection of 183 threshold crossing events with SDEs greater than 10.

Of those 183 candidates, 29 were voted as planet candidates by means of our vetting process; 24 of these are new detections. Our planet candidates have orbital periods ranging from 1.25 -- 6.84 days, and planet radii from 1.26 -- 5.31 \RE. We then discussed the context of these planet candidates in
regards to the radius valley for M-dwarf stars, finding that many of our candidates exist near this planet radius–period domain. With the addition of our new planet candidate detections, along with previous detections observed in sectors 1--5, we calculate an integrated occurrence rate of 2.49 $\pm$ 1.58 planets per star for the period range $\in$ [1,9] days and planet radius range $\in$ [0.5,11] \RE. We project an estimated yield of 122 $\pm$ 11 transit detections of nearby M-dwarfs for the two-year TESS mission. Twenty-three of our new candidates have S/Ns $>$7, Transmission Spectroscopy Metrics $>$38 and Emission Spectroscopy Metrics $>$10, making them promising
targets for follow-up characterization.

As shown in this work, our pipeline is able to produce transit detections for M-dwarf stars with 30-minute cadence light curves, and provides a sample of hidden gems remaining to be found in the FFI data for other sectors observed by TESS. The results of this work are also a testament to the high quality of the FFIs, and invoke a sense of excitement as to other possible worlds that may lie hidden within the data.

All data products for our planet candidates will be released through the Filtergraph visualization portal at \textsf{Filtergraph} visualization portal at \url{https://filtergraph.com/NEMESIS}.\\

\textit{Software Used:} This research made use of \textsf{Python} \citep{Python}, \textsf{Astroquery} \citep{Astroquery:2019}, \textsf{TESSCut} \citep{Brasseur:2019}, \textsf{Transit Least Squares} \citep{Hippke:2019}, \textsf{Scipy} \citep{Scipy}, \textsf{Numpy} \citep{Numpy}, \textsf{Matplotlib} \citep{Matplotlib}, \textsf{Astropy} \citep{Astropy:2018}, \textsf{W\={o}tan} \citep{Wotan}, \textsf{exoplanet} \citep{exoplanet:exoplanet} and \textsf{exoplanet's}
dependencies \citep{exoplanet:kipping13, exoplanet:pymc3, exoplanet:theano, exoplanet:luger18, exoplanet:agol19}, \textsf{EDI-Vetter Unplugged}$^{\ref{ftn:edi_vetter_unplugged}}$ \citep{Zink:2020}.

This paper includes data collected by the TESS mission, obtained from the MAST data archive at the Space Telescope Science Institute (STScI). Funding for the TESS mission is provided by the NASA Explorer Program. STScI is operated by the Association of Universities for Research in Astronomy, Inc., under NASA contract NAS 5-26555. This research has made use of the Exoplanet Follow-up Observation Program website, which is operated by the California Institute of Technology, under contract with NASA under the Exoplanet Exploration Program. The Digitized Sky Surveys were produced at the Space Telescope Science Institute, under U.S. Government grant NAG W-2166. The images of these surveys are based on photographic data obtained using the Oschin Schmidt Telescope on Palomar Mountain, and the UK Schmidt Telescope. The plates were processed into their present compressed digital form with the permission of these institutions. This work was conducted in part using the resources of the Advanced Computing Center for Research and Education at Vanderbilt University, Nashville, TN. This research has also made use of the ADS bibliographic service, and the \textsf{Filtergraph} data visualization service \citep{Burger:2013} at \url{https://filtergraph.com}.

Work by D.L.F. was funded by NASA grant 17-XRP17 2-0024. D.L.F. and K.G.S. acknowledge support from NASA XRP grant 80NSSC18K0445. P.P. acknowledges support from NASA (awards 80NSSC20K0251 and 80NSSC21K0349, and the Exoplanet Exploration Program), the National Science Foundation (Astronomy and Astrophysics grants 1716202 and 2006517), and the Mount Cuba Astronomical Foundation and George Mason University start-up funds. B.A.V. acknowledges support from the Central American–Caribbean Bridge in Astrophysics remote-REU program. The authors acknowledge the guidance and valuable input of Drs. Joshua Pepper, Karen Collins, Joey Rodriguez, Michael Lund, and Avi Shporer. The authors extend special thanks to Dr. Jon Zink, who developed the \textsf{EDI-Vetter Unplugged} software at our request. The authors also thank the anonymous referee for helpful suggestions, which improved this manuscript. \\

\section*{Appendix}
\renewcommand{\thesubsection}{\Alph{subsection}}

\subsection{NEMESIS Validation Reports}\label{sec:appendix_validation_reports}
In this section, we describe our one-page validation reports used in the planet candidate vetting process as described in Sections \ref{sec:vetting} and \ref{sec:EDI}.

\begin{figure*}[htp!]
    \includegraphics[width=0.95\textwidth,height=0.5\textheight, trim=0cm 0.0cm 0cm 0.0cm, clip=true]{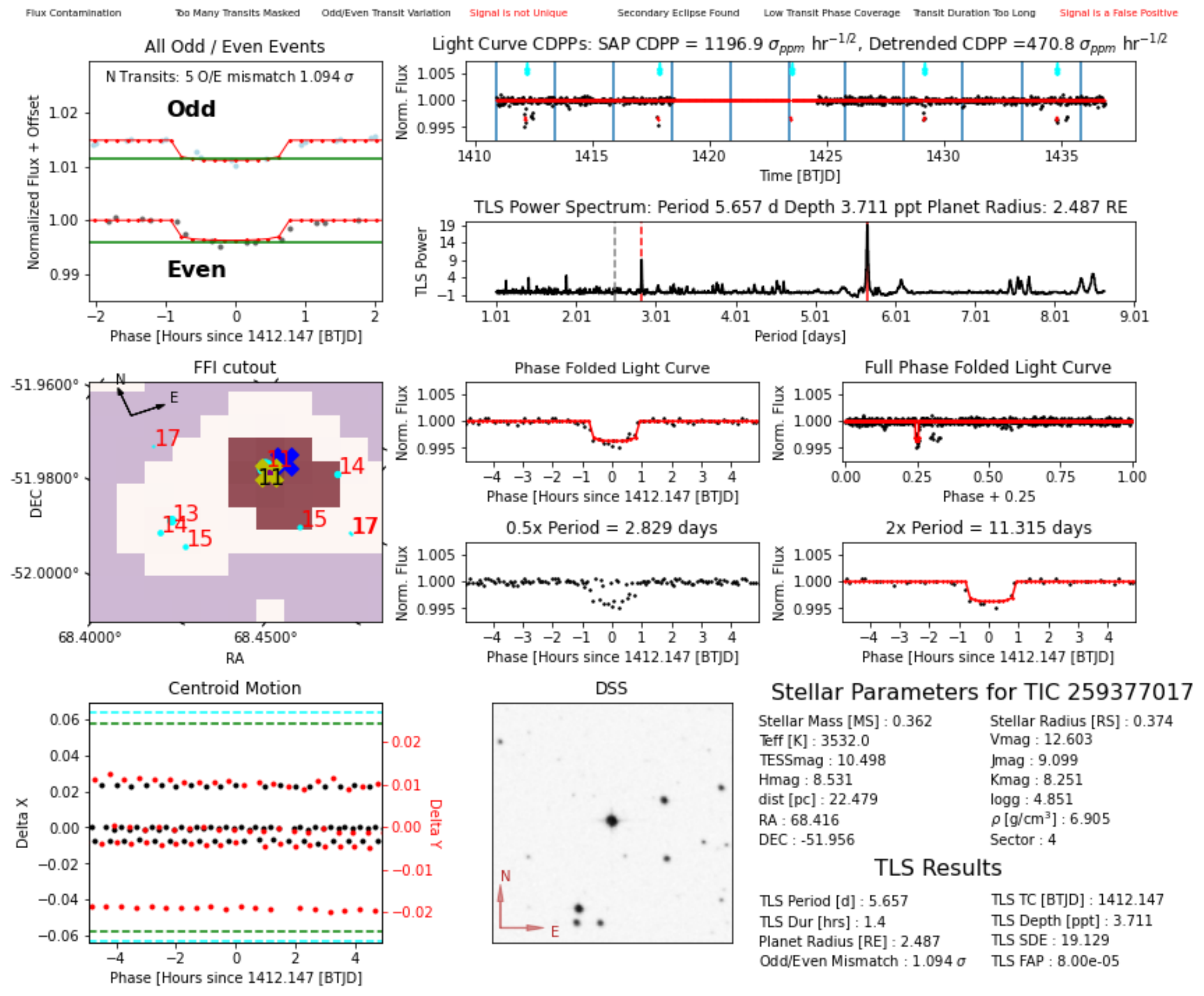}
    \caption{Example of NEMESIS TLS Validation Report for the transiting planet TOI 270 c (TIC 259377017). Top left: odd–even mismatch test; we take the oddand even-numbered transits, and compare the value of each transit depth, while visually verifying that transit duration and shape are consistent. We also display the best-fit TLS transit model in red. Middle left: FFI image cutout; an 11$\times$11 pixel cutout of the FFI. We display the target star as a blue X, with the center of the centroid of as a yellow X, and the selected aperture mask in red, with the background shown in purple. Nearby stars are labeled with their Gaia magnitude via cyan circles and red text. To visually check whether the centroids are on-target, we also place small purple dots marking the centroid positions at each cadence. Lower left: centroid motion; to help keep track of centroid positions, we also plot the motion in the image column (black points) and image row (red points) pixel positions in time, before phase folding the timestamps relative to the TLS-detected transit time and period. Upper right: light curve; we display the full light curve in black points, with the best-fit TLS model in red. The transit times are marked by cyan arrows, and the momentum dumps are marked by vertical blue lines. We also display the CDPP noise metrics for the light curve before processing for the SAP light curve, and afterwards for the detrended light curve. Middle right: TLS power spectrum; we display the TLS power spectra in terms of the SDE as a function of orbital period, measured in days. We mark the momentum-dump rate with a vertical gray dashed line; the TLS period corresponding to the strongest SDE peak is marked with a solid red line, and the 0.5x and 2x harmonics of the TLS period are marked with dashed red lines. Middle: phase-folded light curves; we fold our light curve centered on the transit time, for the full orbital phase and at 0.5, 1, and 2 times the TLS period. Lower-middle right: DSS cutout; to help visualize our target star we also query the Digitized Sky Survey (DSS) catalog, which has a smaller pixel scale than TESS, for quick reference. Lower right text: we display various stellar parameters queried from the TIC and best-fit planet parameters output by TLS. Top center text: EDI-Vetter Unplugged; using a modified version of the EDI-Vetter tool (see footnote $\ref{ftn:edi_vetter_unplugged}$), which uses output from TLS, we conduct several automated planet candidate vetting tests, with false-positive results marked in red text, suggesting that a candidate requires closer inspection}.
    \label{fig:TLS_Report_example}
\end{figure*}

Within the validation reports, we display the light curve in various formats to provide a visual aid to identify the transits detected by TLS. As shown in the upper-right panel of Figure \ref{fig:TLS_Report_example}, we display the light curve in black points, with the best-fit TLS model in red. We also mark the transit times with cyan arrows, and the momentum-dump times with vertical blue lines. We display the CDPP noise metrics for the light curve, before and after processing, for the SAP light curve and the detrended light curve. In the middle-right panel, we display the TLS power spectrum, where the strongest peak is marked by a solid red vertical line, aliases are marked by dashed red vertical lines, and the momentum-dump rate for the observed TESS sector is marked by a dashed gray vertical line. In the middle panels, we display the light curves folded in phase and zoomed in on the transit times using 1/2, 1, and 2 times the TLS-detected period. This folding test is to help determine whether transits folded on 1/2 or 2 times the TLS period appear visually more convincing, and to aid in the detection of eclipsing binaries, where the primary or secondary transits may display different depths. For each validation report, we display the FFI with our selected aperture, and background masks colored in red and purple, as shown in the bottom left panel of Figure \ref{fig:TLS_Report_example}. For each FFI cutout, we also display other nearby stars by referencing Gaia DR2 (\citealt{Gaia:2017}, \citealt{Gaia:2018}), marking their pixel positions with cyan circles, and their Gaia magnitudes in red text. For comparison, we also utilize archival images from the Digitized Sky Survey (DSS), where the image cutouts cover the same region of sky as the TESS FFIs, and are centered on the target star. The pixel scale ofthe DSS is approximately 1.7\arcsec per pixell, and makes visual confirmation of nearby stars easier to determine. Using the best-fit transit parameters initially output by TLS, we display an example of our vetting reports in Figure \ref{fig:TLS_Report_example} for the planet TOI-270 c (TIC 259377017), observed in Sector~4.

\subsection{Planet candidates detected and missed from the TOI catalog}\label{sec:appendix_comparison} 
\begin{figure*}[htp]
    \includegraphics[width=0.95\textwidth]{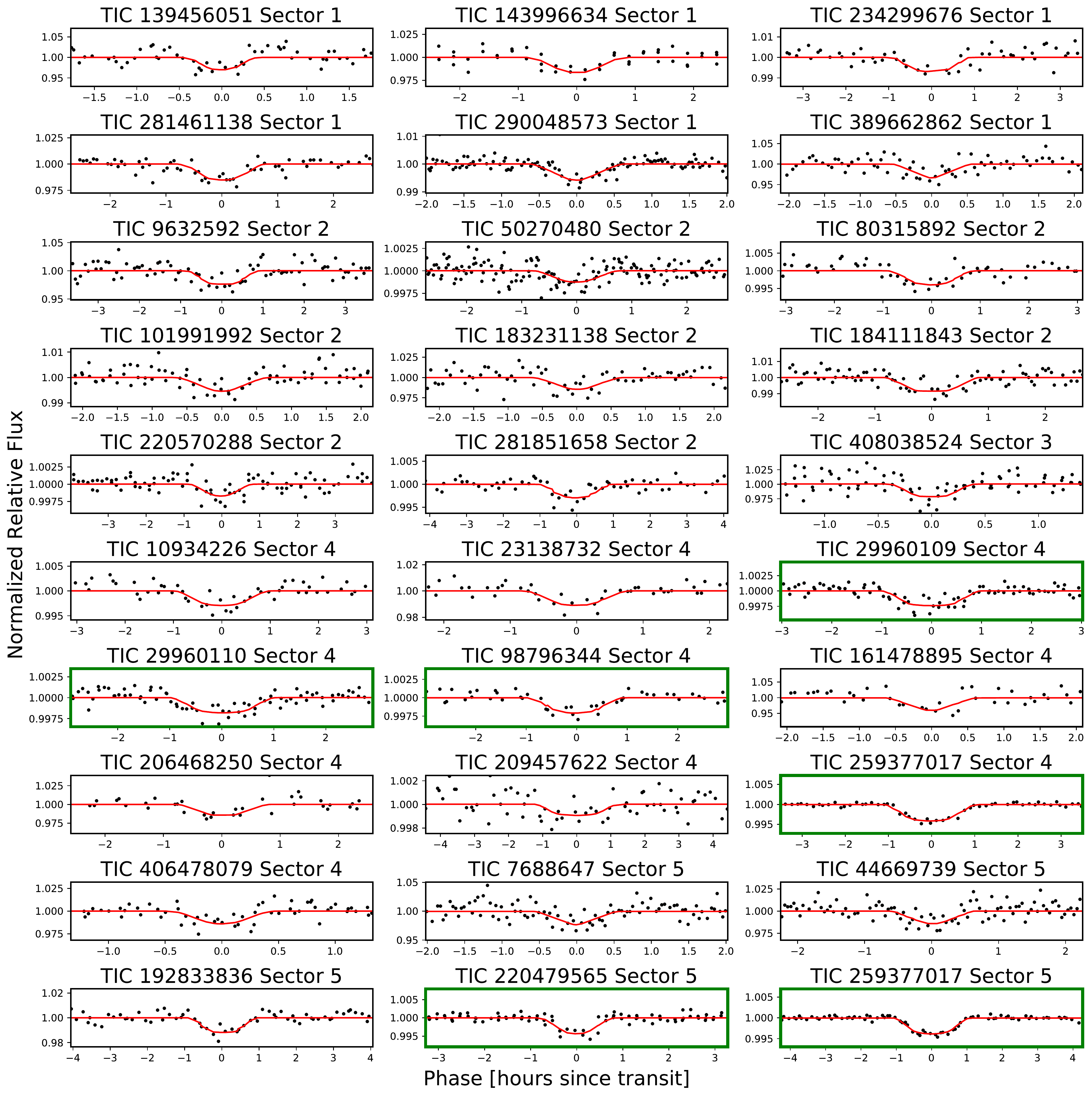} 
    \caption{A gallery of our 29 planet candidate phase-folded light curves (black points). The red lines are the median transit models from our MCMC analysis. The median of the posteriors for the transit model parameters for each candidate can be found in Table \ref{tbl:MCMC_results}. Planet Candidates outlined by green borders are TOIs.}
    \label{fig:PC_summary}
\end{figure*}

In this section, we discuss the planet candidates from the TOI catalog that were missed and detected by the NEMESIS Pipeline. We also present a gallery of our 29 planet candidates in Figure \ref{fig:PC_summary}.

TIC 29960109 (TOI 393) and TIC 29960110 (TOI 1201):
based on the observation notes listed on the Exoplanet Follow-up
Observing Program for TESS (ExoFOP-TESS) website\footnote{https://exofop.ipac.caltech.edu/tess}, 393 is listed as a nearby eclipsing binary (NEB), while TOI 1201 remains an active planet candidate. TOIs 393 and 1201 are located on the same TESS pixel, and follow-up observations indicate that the transit signal is likely to originate from TOI 1201.

TIC 98796344 (TOI 455) is a multistar system, where three mid-to-late M-dwarfs lie on a single TESS pixel. The planet, discovered by \citealt{Winters:2019} has a radius of 1.38 \RE~and an orbital period of 5.35882 days. As shown in Table \ref{tbl:MCMC_results}, our initial TLS detection has an accurate match for the orbital period, and a planet radius of 1.3391 \RE. Our MCMC analysis also provided a comparable radius of 1.26 \RE which is within the published uncertainty, and makes this a successful detection. 

\textit{TIC 259377017 (TOI 270):} TOI 270 is a three planet system, discovered by \citealt{Gunther:2019}, that consists of TOI 270 b (R$\mathrm{_P}$ = 1.247 \RE, P = 3.36 days), TOI 270 c (R$\mathrm{_P}$ = 2.42 \RE, P = 5.66 days) and TOI 270 d (R$\mathrm{_P}$ = 2.13 \RE, P = 11.38 days). The planets orbit close to a mean-motion resonant chain, where TOIs 270 b and c have a period ratio of 5:3, and TOIs 270 c and d have a period ratio of 2:1. The TLS power spectra of our initial detection of TIC 259377017 exhibits strong power at both the 3.36- and 5.66-day periods, with the 5.66-day signal having more power, as it is a largertransit. Due to our transit search strategy of only considering the strongest peak in the TLS power spectra, and searching only up to nine days, we did not automatically detect TOI 270 b and TOI 270 d with our pipeline, although other transits are visually present, as shown in the upper-right panel of Figure \ref{fig:TLS_Report_example}. As for TOI 270 c, we detected the planet in both sectors four and five. TOI 270 c was also observed in sector three, but was missed by our pipeline, due to particularly high scatter for that light curve. Our initial TLS-measured planet radii closely match the published results; our MCMC radii for both sectors are slightly undersized, but are still within the published uncertainty.

TIC 220479565 (TOI 269) is an active planet candidate, initially detected by the SPOC pipeline, with R$_{P}$ = 2.9519 \RE, and P = 3.698 days which was observed via a two-minute cadence. We detected TOI 269 in sector five, with a close-matching period and radius with respect to our MCMC analysis. TOI 269 was also observed in sectors three and four, but our pipeline missed those transit detections. Our sector three and four SAP light curves have significant instrumental effects near the beginning and end of the satellite orbits. Our PLD, smoothing, and outlier rejection processes were able to reduce the light curve CDPP from 2940 to 2440 ppm in sector three, and 4100 ppm to 800 ppm in sector four, but regions of higher-than-average scatter still remained, which may have interfered in the automated detection of the transits.

TIC 200322593 (TOI 540) hosts a small, short-period planet (RP = 0.9 \RE, P = 1.239 days), orbiting a rapidly rotating M-dwarf (P$_{rot}$ = 17.43 days), discovered by \citealt{Ment:2020}. We detected a transit signal with R$\mathrm{_P}$ = 0.767 \RE and P= 2.488 days, which is about twice the published period, but ultimately did not appear transit-like in our validation report, and was voted a false positive. With the automated nature of our pipeline, our smoothing window (for this star, about 4.86 hours) is possibly under-optimized for fully removing the stellar rotation signal.

Both TIC 231702397 (TOI 122) and TIC 305048087 (TOI
237) were missed by our pipeline, but are both published planets, discovered by \citealt{Waalkes:2020}. Our pipeline detected a transit event for TOI 122 with R$_{P}$ = 2 \RE, P = 8.37 days, which was vetted as a false positive, and differs from the published values of R$_{P}$ = 2.72 \RE, P = 5.078 days. For TOI 237, our pipeline detected a transit event with R$_{P}$ = 1.4 \RE, P = 8.03 days (which also differs from the published values of R$_{P}$ = 1.44 \RE, P = 5.436 days), and was voted as a planet candidate; however, due to its SDE of 8.4, it was excluded from this work.

There are a handful of other known planet- and planet candidate-hosting stars that were missed by our pipeline. TIC 150428135 (TOIs 700 b, c d), TIC 410153553 (TOI 136 b; LHS 3844 b), TIC 12822545 (K2-54 b), TIC 92226327 (TOI 256 b and c; LHS 1140 b and c), TIC 415969908 (TOI 233), TIC 12421862 (TOI 198), TIC 153065527 (TOI 406), TIC 153077621 (TOI 454), TIC 259962054 (TOI 203), and TIC 77156829 (TOI 696) were all missed by our pipeline, due to their published/catalog periods being either too short ($<$1 day), or too long ($>$9 days).

\bibliographystyle{apj}
\clearpage
\bibliography{main}

\end{document}